\documentclass[modern]{aastex631}

\newcommand{\ecss}{erg~cm$^{-2}$~s$^{-1}$~sr$^{-1}$}
\newcommand{\lam}{$\lambda$}
\newcommand{\kms}{km~s$^{-1}$}
\renewcommand{\ion}[2]{#1\,{\sc #2}}
\newcommand{\as}{$^{\prime\prime}$}
\newcommand{\hinode}{\textit{Hinode}}
\newcommand{\pry}[1]{#1}

\received{June 1, 2019}
\revised{January 10, 2019}
\accepted{\today}
\submitjournal{ApJ}

\graphicspath{{./}{figures/}}
\begin{document}

\title{Scattered light in the \hinode/EIS and SDO/AIA instruments measured from the 2012 Venus transit}

\author[0000-0001-9034-2925]{Peter R. Young}
\affiliation{NASA Goddard Space Flight Center \\
Greenbelt, MD 20771, USA}
\affiliation{Northumbria University, Newcastle upon Tyne, NE1 8ST, UK}

\author[0000-0003-1692-1704]{Nicholeen M. Viall}
\affiliation{NASA Goddard Space Flight Center \\
Greenbelt, MD 20771, USA}

\begin{abstract}
Observations from the 2012 transit of Venus are used to derive empirical formulae for long and short-range scattered light at locations on the solar disk observed by the \textit{Hinode} \textit{Extreme ultraviolet Imaging Spectrometer} (EIS) and the \textit{Solar Dynamics Observatory} \textit{Atmospheric Imaging Assembly} (AIA) instruments. Long-range scattered light comes from the entire solar disk, while short-range scattered light is considered to come from a region within 50\arcsec\ of the region of interest. \pry{The formulae were derived from the \ion{Fe}{xii} 195.12~\AA\ emission line observed by EIS and the AIA 193~\AA\ channel.}
\pry{A study of the weaker \ion{Fe}{xiv} 274.20~\AA\ line during the transit, and a comparison of scattering in the AIA 193~\AA\ and 304~\AA\ channels suggests the EIS scattering formula applies to other emission lines in the EIS wavebands}. \pry{Both formulae should be  valid in regions of fairly uniform emission such as coronal holes and quiet Sun, but not faint areas close (around 100\as) to bright active regions.}
%apply to the 193~\AA\ channel of AIA and to emission lines formed at around 1.5~MK by the EIS instrument\pry{They
The formula for  EIS is used to estimate the scattered light component of  \ion{Fe}{xii} \lam195.12 for seven on-disk coronal holes observed between 2010 and 2018. Scattered light contributions of \pry{56}\%\ to 100\%\ are found, suggesting that these features are dominated by scattered light, consistent with earlier work of Wendeln \& Landi.  \pry{Emission lines from the \ion{S}{x} and \ion{Si}{x} ions---formed at the same temperature as \ion{Fe}{xii} and often used to derive the first ionization potential (FIP) bias from EIS data---are also expected to be dominated by scattered light in coronal holes.}
\end{abstract}

\section{Introduction} \label{sec:intro}

Images of the solar corona obtained in extreme ultraviolet (EUV) emission lines broadly separate by brightness into dark coronal holes, quiet Sun, and bright active regions.
Within these features, emission is highly inhomogeneous with compact bright points, and extended loops and plume structures. When interpreting the image data, considerations of scattered light (often referred to as stray light) within the instrument can be important. For example, bright features have their intensities reduced due to scattering, while dark areas can be contaminated by scattered light from neighboring bright areas. EUV emission is optically thin and thus emitted photons are not scattered within the corona---the scattering is entirely within the instrument.

The present work was motivated by the need to make spectroscopic measurements 
%in and around coronal holes as part of activities 
to relate coronal plasma properties with those measured in situ. A fundamental objective for Heliophysics is to determine how structures in the solar atmosphere produce structure in the solar wind,
%how solar wind structures evolve from structures in the solar atmosphere, 
which is particularly relevant to the Parker Solar Probe \citep{2016SSRv..204....7F} and Solar Orbiter \citep{2020A&A...642A...1M} missions launched in 2018 and 2020 that will reach heliocentric distances of  0.05~AU and 0.28~AU, respectively. By making in situ measurements in the outer solar corona and young solar wind \citep{2020JGRA..12526005V} 
%close to the solar corona 
there is a greater possibility of measuring plasma that is relatively unevolved from its coronal origins. 

Coronal holes are locations where some of the photospheric magnetic field opens directly into the heliosphere and their centers are widely acknowledged as the source regions of the fast solar wind \citep{1976SoPh...49..271S}.
The boundary regions of coronal holes are also believed to make a contribution to the more variable, slower solar wind, either by direct release of plasma along the open magnetic field lines \citep[e.g.,][]{1990ApJ...355..726W}, or indirectly through interchange reconnection of open field lines with the closed field of the nearby quiet Sun and/or active regions \citep[e.g.,][]{2001ApJ...560..425F,2011ApJ...731..112A}. 
EUV spectrometers are important for measuring parameters such as velocity (via Doppler shifts), temperature and element abundances in coronal holes that can be related to the parameters measured in situ. Due to the low coronal intensities within coronal holes it is thus vital that the scattered light component to coronal hole measurements be quantified.

Ideally scattered light should be characterized before the launch of a space instrument, but technical limitations or time constraints can prevent this. 
This was the case for the \textit{EUV Imaging Spectrometer} \citep[EIS;][]{culhane07} on board the \hinode\  spacecraft \citep{2007SoPh..243....3K}, which obtains spectral images in the 170--212 and 246--292~\AA\ wavelength regions. \pry{Each emission line observed by EIS has a characteristic temperature of formation [$T_\mathrm{f}$] for which the ion has its peak abundance in equilibrium conditions. These temperatures are obtained from the CHIANTI atomic database \citep{2016JPhB...49g4009Y,2021ApJ...909...38D}.} 
For observations above the solar limb, the scattered light can be estimated by measuring emission lines with $\log\,(T_\mathrm{f}/\mathrm{K})\le5.5$ as these are expected to have zero intensity at coronal heights \citep{2011ApJ...736..101H}. This method does not work for on-disk measurements, however. 

The only study of on-disk scattered light is by \citet{2018ApJ...856...28W}, who measured the intensities of EIS emission lines with different values of $T_{\rm f}$, inside and outside of two on-disk coronal holes. They found that the coronal hole to quiet Sun ratios decreased with \pry{increasing} temperature until $\log\,(T_{\rm f}/\mathrm{K})=6.15$, above which the ratios were constant.
Given that coronal holes are known to be cooler than quiet Sun regions \citep[e.g.,][]{1998A&A...336L..90D}, the authors concluded that the on-disk coronal hole intensities are dominated by scattered light for ions with $\log\,(T_{\rm f}/{\rm K})\ge 6.15$.
%, and thus the measured intensities actually represented the surrounding quiet Sun plasma rather than the coronal hole.
\pry{This result does not imply that the behavior of scattered light is changing as a function of temperature. All emission lines in the coronal hole will have a scattered light  component coming from the surroundings. Rather, this result arises because the coronal hole signal becomes increasingly weak at higher temperatures and thus eventually becomes overwhelmed by the scattered light component.}

\pry{The \citet{2018ApJ...856...28W} result implies, in particular,} that emission lines of \ion{Si}{x} and \ion{S}{x}, both with $\log\,(T_{\rm f}/{\rm K})=6.15$, are predominantly composed of scattered light. These two ions are used to infer the Si/S element abundance ratio \citep{2009ApJ...695...36F,2011ApJ...727L..13B}, which is used as a diagnostic of the \textit{FIP bias}, i.e., the enhancement factor of elements with low first ionization potential (FIP) often found in coronal and solar wind plasma. The FIP bias is determined by processes in the chromosphere and low corona; once a parcel of plasma is released into the heliosphere, its FIP bias is frozen in, and does not evolve as the solar wind advects outward. Thus the FIP bias is a very important measurement for connecting structures and variability in the solar wind with their coronal source \citep{1998SSRv...85..253P,2015LRSP...12....2L}. The \citet{2018ApJ...856...28W} results thus suggest that efforts to use the Si/S FIP bias ratio in coronal holes to find connections to the solar wind plasma measured by Solar Orbiter are likely to have large uncertainties.

The \citet{2018ApJ...856...28W} method requires an intensity comparison between the coronal hole and surrounding quiet Sun across a range of ions formed at different temperatures. It does not intrinsically yield an estimate of the scattered light, but instead identifies a break point in the coronal hole/quiet Sun ratio vs.\ temperature relation where the ratio becomes constant, implying scattered light is completely dominant. An additional complication is that, due to the generally small size of the EIS rasters, the quiet Sun measurement is made close to the coronal hole. If the scattered light in the coronal hole is coming from larger distances, then the local quiet Sun emission  may not accurately reflect the full extent of the scattered light.

In the present work we use the 2012 transit of Venus to obtain a general purpose empirical formula for scattered light in EIS observations using \ion{Fe}{xii} emission observed by EIS and the \textit{Atmospheric Imaging Assembly} \citep[AIA:][]{2012SoPh..275...17L} on the \textit{Solar Dynamics Observatory} \citep[SDO:][]{2012SoPh..275....3P}. In comparison to the \citet{2018ApJ...856...28W} work, the formula yields a direct percentage estimate of scattered light at any location on the solar disk. By cross-calibrating quiet Sun intensities between EIS and AIA, the full-disk intensity from AIA is used to estimate long-range scattering. The formula is derived from data obtained with the 40\as\ slit (or ``slot") of EIS, and the companion work \citet{2022arXiv220314161Y} provides technical details of EIS slot data, including a correction factor for measured intensities that is applied here.
\pry{We argue in Section~\ref{sect.ext} that the formula can be used for any line within the EIS wavelength ranges, and} it should be valuable in assessing whether a specific EIS coronal hole observation exhibits a significant degree of scattered light.

Transits of Venus across the face of the Sun are rare celestial events, with two occurring eight years apart and the preceding and following pairs more than 100 years apart. The most recent pair occurred in 2004 and 2012, and the next transit will not be until 2117. The  Venus shadow as it appears against the solar disk has an angular diameter of 60\arcsec, which is sufficiently large that short-range scattered light is significantly reduced in the center of the shadow, but not so large that the scattered light from the full solar disk is reduced. Hence the transit is valuable for assessing the relative contributions of short and long-range scattered light. This is in contrast to the much more frequent Mercury transits (angular diameter 10\arcsec) and partial solar eclipses by the Moon. A large portion of the Sun is blocked in the latter, and the lunar limb moves so quickly (a few arcsec per second) that it is blurred in typical EIS exposures of tens of seconds, preventing  measurements of short-range scattered light.

% Our article focuses on the EIS \ion{Fe}{xii} \lam195.12 emission line that is formed at $\log\,(T_\mathrm{f}/\mathrm{K})=6.15$ and is usually the strongest line (in terms of photon counts) measured by the instrument. The formation temperature is very similar to that of \ion{Si}{x} and \ion{S}{x}, the ions most commonly used for FIP bias measurements, and so we argue that the results extend to these ions. The AIA 193~\AA\ channel in typical solar conditions is dominated by \ion{Fe}{xii} (including the 195.12~\AA\ line) and, to a lesser extent, \ion{Fe}{xi} \citep{2010A&A...521A..21O}. This will allow us to use the  AIA 193~\AA\ intensity averaged over the full-disk as a proxy for the full-disk averaged \lam195.12 intensity, which is not available from EIS.

Section~\ref{sect.scatt} discusses previous studies of scattered light in EUV instruments, and Section~\ref{sect.soft} summarizes the analysis software used in this article. Section~\ref{sect.aia} presents an analysis of the AIA data obtained during the transit, yielding a general formula for scattered light in the instrument. Section~\ref{sect.eis} describes the EIS transit observations, and the analysis of Section~\ref{sect.eis-anal} yields an empirical formula describing the scattered light. Section~\ref{sect.pres} gives our prescription for how to derive the scattered light contribution for on-disk EIS measurements of the \ion{Fe}{xii} \lam195.12 line. Examples from seven coronal hole observations are given in Section~\ref{sect.ch}. \pry{In Section~\ref{sect.ext} we consider whether the scattering formula for \ion{Fe}{xii} \lam195.12 applies to other emission lines in the EIS wavelength bands.} Our results are summarized in Section~\ref{sect.summary}, and a discussion of their significance is given in relation to FIP bias and Dopper shift measurements in coronal holes.

%-----------------------------
\section{Scattered light for EUV instruments}\label{sect.scatt}

In simple terms, the scattered light for an on-disk solar observation can be considered to consist of three components: local, short-range and long-range. To illustrate the difference, consider a dark, circular coronal hole of radius 100\as. In the center there is a tiny, intense bright point that is around a factor 100 brighter than the coronal hole. Outside the coronal hole the quiet Sun emission is completely uniform and around a factor 10 brighter than the coronal hole. The point spread function (PSF) that describes how the light from a point source is spatially dispersed across a detector typically consists of a narrow core and broad wings. If the instrument resolution is 2\as, then the core can be considered a Gaussian of full-width at half-maximum of 2\as. The emission from the bright point due to this core will extend out to around 4\as\ due to the contrast between the coronal hole and the bright point. The wing emission may extend further to 5--10\as. This is local scattered light. 

The quiet Sun emission around the coronal hole is weaker but is distributed continously. A block of $3\times 3$ pixels will yield similar wing emission to the bright point, but the combined effect of multiple nearby blocks of pixels will serve to create a more significant wing component than the bright point. This can be expected to extend 10's of arcsec into the coronal hole, but diminish towards the coronal hole center. This is short-range scattered light.

Finally, the presence of huge numbers of quiet Sun pixels from the entire solar disk, but far from the coronal hole, will result in a scattered light signal in the coronal hole that is due to the very faint far wings of the PSF. This is long-range scattered light and would be expected to be constant across the coronal hole.

\pry{The PSF described above is a function that varies smoothly from the core to the wings, and radially (or close-to radially) symmetric.}
%\pry{For EUV instruments an optical component that makes an important contribution to the PSF is a mesh filter placed before the primary mirror to block visible light.}
A complication for EUV instruments such as AIA and EIS is the presence of mesh filters in the optical path that are used to block visible light. A compact source such as the coronal hole bright point discussed above will produce a complex diffraction pattern that, to first order, appears as a cross or double-cross  on the detector. Examples from EIS can be seen in Figure~9 of \citet{2018ApJ...856...28W} and an example from AIA is shown in Figure~1 of \citet{2011ApJ...743L..27R}. A model of the AIA PSF is shown in Figure~5 of \citet{aia-psf}. The scattered light in the coronal hole would thus be enhanced at the cross locations, but not otherwise. For the continuous quiet Sun emission, the effect of the filter diffraction pattern is to produce an enhanced, smooth, symmetric wing to the scattered light.

Scattered light in AIA data has been studied by previous authors, building on earlier work
with the predecessor \textit{Transition Region And Coronal Explorer} \citep[TRACE;][]{1999SoPh..187..229H} mission, and we briefly summarize these activities here.

\citet{2001SoPh..198..385L} used flare data obtained with TRACE to study the diffraction pattern from the mesh filter, finding that the pattern contains around 20\%\ of the incident light. They also found that the dispersion within the individual diffraction orders is sufficient to resolve spectral features, and this latter feature has been used for diagnostic purposes by \citet{2011ApJ...734...34K} and \citet{2011ApJ...743L..27R} using TRACE and AIA data, respectively. Tens of diffraction orders can be seen in large flares and they can be used to derive flare intensities in the case that the zeroth order image is saturated on the detector \citep{2014ApJ...793L..23S}.
The diffraction patterns for all of the AIA EUV channels were computed in a technical report by \citet{aia-psf}, and are implemented in IDL software available in the \href{https://sohoftp.nascom.nasa.gov/solarsoft/}{\textsf{SolarSoft}} library \citep{1998SoPh..182..497F,2012ascl.soft08013F}.

\citet{2009ApJ...690.1264D} were the first to model the complete PSF of TRACE by including the diffraction pattern, a narrow, Gaussian core, a broader Gaussian ``shoulder", and an isotropic, long-range component modeled as a Gaussian-truncated Lorentzian. The latter effectively corresponds to the PSF wing discussed earlier. This approach was extended to AIA data by \citet{2013ApJ...765..144P} and has been used by other authors to deconvolve their AIA images \citep[e.g.,][]{2021ApJ...906...62L,2021ApJ...907....1U}. We note that the \citet{2013ApJ...765..144P} PSF puts all of the scattered light within 100\arcsec\ of the source.

\citet{2016JSWSC...6A...1G} proposed a general-purpose, non-parametric blind deconvolution scheme that can be applied to any image data. They applied it to AIA data from the 2012 Venus transit and compared the results with the parametric model of \citet{2013ApJ...765..144P}, finding comparable results. An important point is that deconvolutions with both the non-parametric and parametric PSFs did not lead to zero signal in the Venus shadow and so the authors required an additional constant level of scattered light across the image. The implication is that there is an additional component of scattered light  beyond 100\arcsec\ from the source. A similar finding was made by \citet{2012ApJ...749L...8S} using data from the Extreme Ultraviolet Imager (EUVI) on board STEREO. Our interpretation is that this emission comes from the very far wings of the PSF that were not modeled in these articles.

An imaging telescope with a large field-of-view such as AIA is ideal for investigating the full PSF, including the long-range component. The situation is more complex for an imaging slit spectrometer such as EIS, however. The slit isolates a narrow column of the image focused by the primary mirror. Thus information about the wider image field that gives rise to the short and long-range scattering is lost. The field can be built up by performing a raster scan but at most this will cover around 600\arcsec\ $\times$ 500\arcsec, and usually significantly less due to telemetry or cadence restraints.

Our solution in the present work is to find a simple empirical formula that yields an estimate of scattered light in EIS data at on-disk locations. A comparable formula is first derived using AIA data from the Venus transit to illustrate the method, and then applied to EIS.

\section{Analysis Software}\label{sect.soft}

The analysis performed for this article used IDL code written by the authors or contained in the \textsf{Solarsoft} IDL library. Software that may be generally useful for other EIS or AIA observations have been placed in the \textsf{GitHub} repository \href{https://github.com/pryoung/aia-eis-venus}{\textsf{aia-eis-venus}}. Software specifically created  for the Venus analysis and for generating figures in the present article have been placed in the \textsf{GitHub} repository \href{https://github.com/pryoung/papers/tree/main/2022_venus}{\textsf{pryoung/papers/2022\_venus}}.

Some of the analysis performed here is done with \href{https://hesperia.gsfc.nasa.gov/rhessidatacenter/complementary_data/maps/maps.html}{IDL \textit{maps}}, which place solar images within a common heliocentric coordinate system. This makes it easy to extract co-spatial images from different instruments. AIA maps are created in the present work with the routine \textsf{sdo2map}, and EIS maps are created with \textsf{eis\_slot\_map} and \textsf{eis\_get\_fitdata}. These routines are available in \textsf{Solarsoft}.

%---------------------------------------
\section{AIA Observations and Analysis}\label{sect.aia}

\begin{figure}[t]
    \centering
    \includegraphics[width=\textwidth]{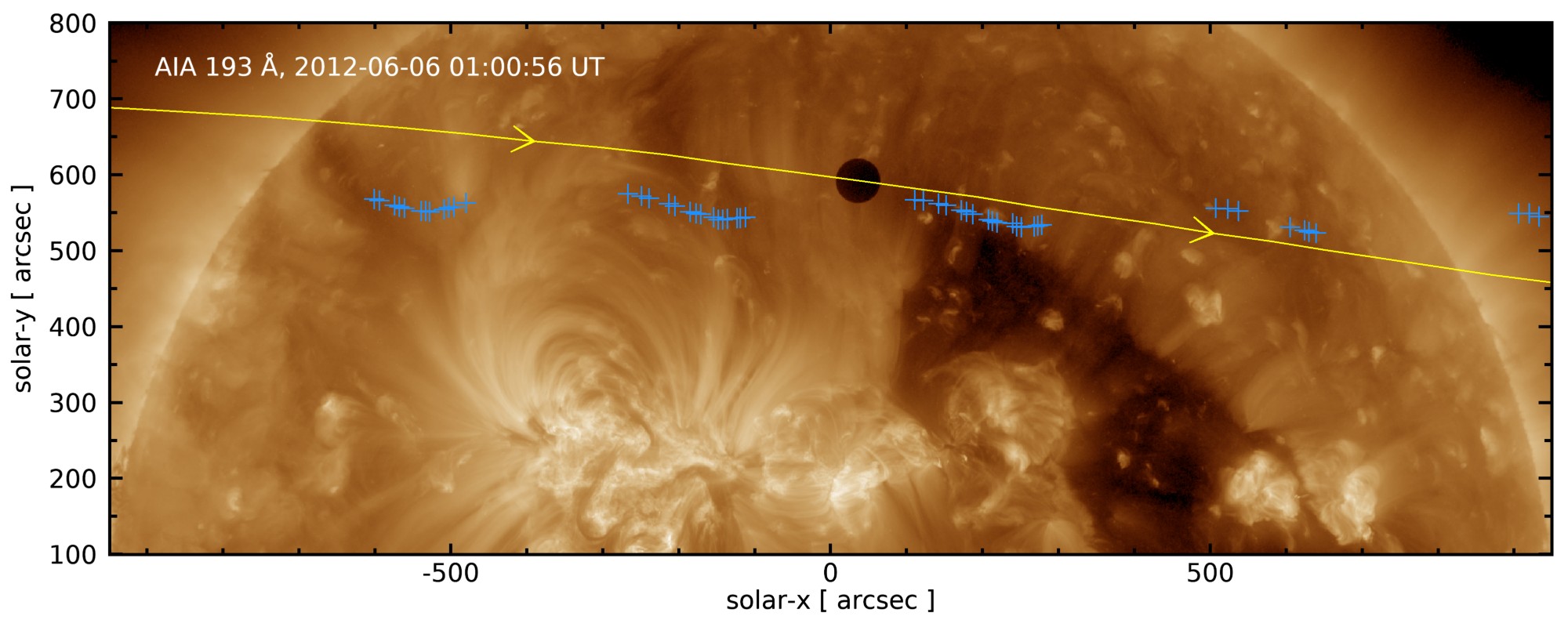}
    \caption{An AIA 193~\AA\ image from 2012 June 6, with Venus located at approximate position [40,590]. The track and direction of motion are indicated by the \textit{yellow line}. A logarithmic intensity scaling has been applied to the image. The \textit{blue crosses} show the locations of Venus as measured by EIS. \pry{The differences compared to the AIA track are due to the different orbits of the \hinode\ and SDO spacecraft (Section~\ref{sect.eis-anal})}.}
    \label{fig.aia-image}
\end{figure}

\pry{AIA obtains full-disk images of the Sun in a number of filters, including seven at EUV wavelengths that are centered on strong emission lines. For the present work, we mostly use images from the 193~\AA\ channel that has dominant contributions from \ion{Fe}{xii} lines at 192.39, 193.51 and 195.12~\AA\ in most conditions. In Section~\ref{sect.ext} we also consider data from the 211~\AA\ channel that is usually dominated by the \ion{Fe}{xiv} 211.32~\AA\ line. The AIA EUV images are obtained at a 12~s cadence and the image pixel size corresponds to 0.6\as.
}

Venus entered the AIA field of view around 20:20~UT on 2012 June 5 and exited around 06:20~UT on June 6. (For the remainder of this article we will drop the dates when specifying times during the transit.) 
The leading edge of Venus was externally tangent to the solar limb (Contact I)  at 22:07~UT and the trailing edge was externally tangent (Contact IV) at 04:37~UT.
Figure~\ref{fig.aia-image} shows an AIA 193~\AA\ image from 01:01~UT, with Venus---visible as a black circle of diameter 60\arcsec---close to the central meridian. 
Venus passed to the north of a large active region complex comprising active regions with numbers 11493, 11496, 11498, 11499 and 11501.  To the west of the complex was a large low-latitude coronal hole, and Venus clipped the northern section of the hole. 

During the transit, C1 and C2-class flares peaked at 23:13~UT and 02:19~UT, respectively, and there were several B-class flares. The background GOES activity was at the B7 level. \pry{The C2-class flare, located near the west limb at ($+822,-220$) produced only a factor three increase in the AIA 193~\AA\ intensity  at that location, and so the impact on the scattered light at Venus is negligible.}
%The C-class flares were around 500\arcsec\ distant from the Venus shadow and thus are not expected to affect the scattered light measured at Venus as they are effectively point sources (see discussion of Section~\ref{sect.scatt}).

The period 21:00~UT to 05:40~UT was selected, corresponding to Venus transiting the corona from  250\arcsec\ above the north-east limb to 200\arcsec\ above the north-west limb. We chose twenty-six AIA 193~\AA\ images spaced at 20-min intervals  for the scattered light analysis, supplemented by six additional images at intervals of 5 or 10~min during periods when the intensity in the Venus shadow changed rapidly. 
%These additional images are indicated by filled circles (above the limb) or diagonal crosses (inside the limb) in Figure~\ref{fig.aia-int}.

For each of the AIA images, the average intensity at the center of the Venus shadow, $D_\mathrm{V}$, and the average intensity for an annulus surrounding the shadow, $D_\mathrm{ann}$, were extracted. The IDL routine \textsf{aia\_get\_venus} was written for this purpose, and is available in the \textsf{2022\_venus} repository. For each of the image frames, the routine displays a close-up of the Venus shadow and the user manually selects the center of the shadow. A box of $33\times 33$ pixels (20\as\ $\times$ 20\as) centered on this location is extracted and averaged to yield $D_\mathrm{V}$. The level-1 AIA files  have had the detector background removed and so there is no need to correct for this. 
\pry{We use the standard deviation of the intensities of the $33\times 33$ pixel block, $\sigma_\mathrm{V}$, as the uncertainty for the Venus intensity measurement.}
The annulus surrounding the shadow has an inner radius of 30\as\ and an outer radius of 50\as. The inner radius is set by the size of the Venus shadow, while the outer radius is set with a view to applying the same method to EIS, which generally has small fields of view. The effects of choosing different radii are considered in Appendix~\ref{app.annulus}, where it is found that the differences would be mostly a few percent, or up to 26\%\ in the worst case. The results from \textsf{aia\_get\_venus} are stored in an IDL save file available in the \textsf{2022\_venus} repository.

\begin{figure}[t]
    \centering
    \includegraphics[width=\textwidth]{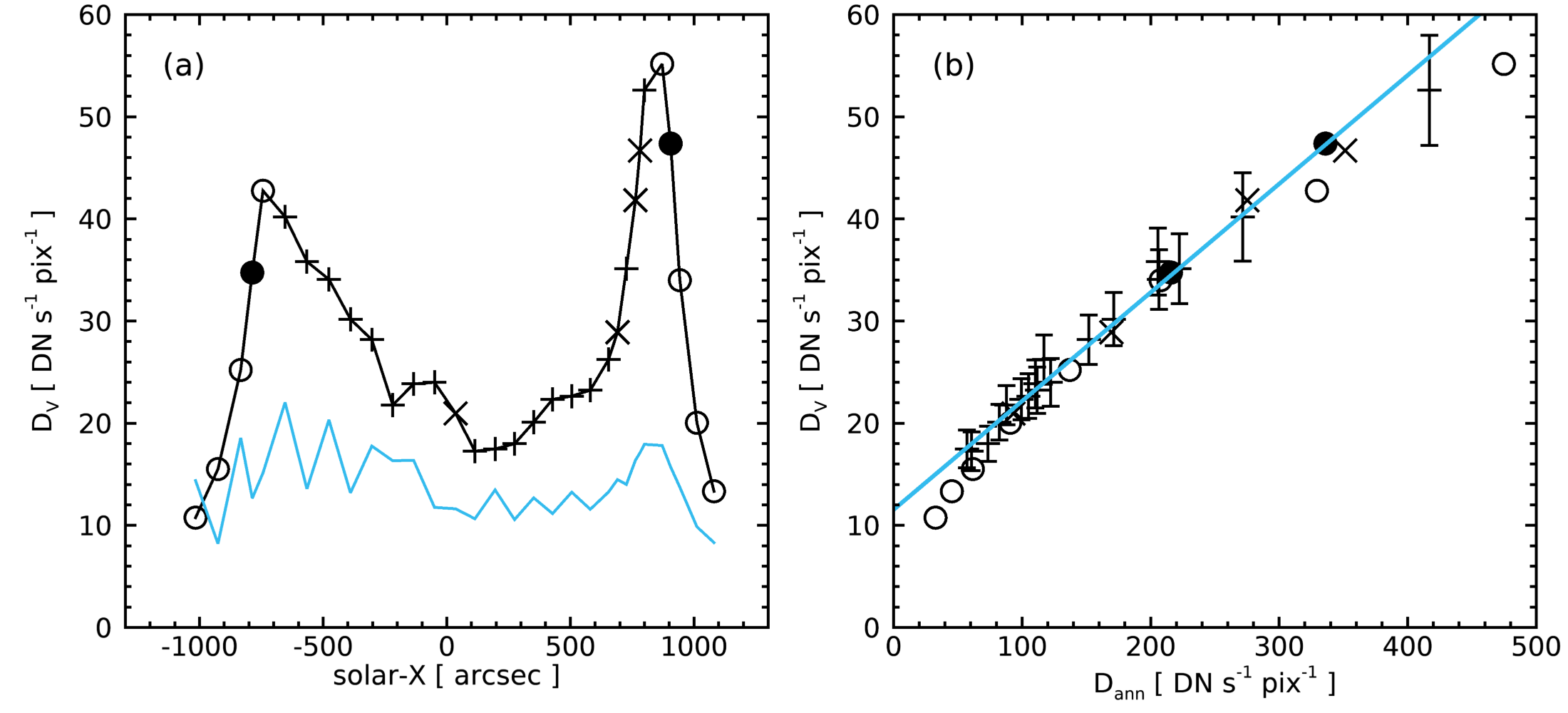}
\caption{Panel (a) plots the Venus intensity, $D_{\rm V}$ as a function of the solar-$x$ position of Venus. \textit{Circles} indicate positions above the limb, and \textit{crosses} positions on the disk. \textit{Filled circles} and \textit{diagonal} crosses are additional points that were not part of the 20-min cadence image series (see main text). The \textit{cyan line} is the Venus intensity after deconvolving the images. Panel (b) plots $D_{\rm V}$ against the annular intensity, $D_{\rm ann}$. Symbols are the same as for panel (a). The \textit{cyan line} is a straight line fit to only the crossed data points.}
\label{fig.aia-int}
\end{figure}

Figure~\ref{fig.aia-int}a plots  $D_{\rm V}$ against solar-$x$ position. The six additional images mentioned above are plotted as filled circles or diagonal crosses. A distinctive horned profile is seen with the two peaks corresponding to the passage of Venus in front of the solar limb. The passage over the west limb occurred at $y=470$\arcsec, about 200\arcsec\ lower than at the east limb, and the coronal emission was brighter here, explaining the larger peak at the west limb.

Also shown in Figure~\ref{fig.aia-int}a is the Venus intensity after the AIA images have been deconvolved with an estimate of the instrument point spread function (PSF). We followed the procedure of \citet{aia-psf}, using the IDL routine \textsf{aia\_psf\_calc} to create the PSF function, and then the routine \textsf{aia\_deconvolve\_richardsonlucy} to deconvolve the images. As noted by \citet{2020SoPh..295....6S}, there is a residual signal in the Venus shadow. Here we find that this signal is approximately constant during the Venus transit so the deconvolution is effective in removing the local scattered light.

Figure~\ref{fig.aia-int}b plots $D_{\rm V}$ against $D_{\rm ann}$, which demonstrates a tight correlation between the AIA intensity immediately adjacent to the Venus shadow and the scattered light within the shadow. A straight line is fit to the points from inside the limb (indicated by crosses in Figure~\ref{fig.aia-int}). The off-limb points were omitted because the $D_{\rm V}$--$D_{\rm ann}$ relation is intended to be applied to on-disk locations, and there is a suggestion from Figure~\ref{fig.aia-int}b that the off-limb points follow a different pattern from the on-disk points.  The slope of the linear fit is $0.1063\pm 0.0077$ and $D_{\rm V}=11.5\pm 1.1$~DN~s$^{-1}$~pix$^{-1}$ for $D_{\rm ann}=0$~DN~s$^{-1}$~pix$^{-1}$. 

The interpretation of the $D_{\rm V}$--$D_{\rm ann}$ relation is that the scattered light within the Venus shadow scales linearly with the local emission, but with a constant background due to the long-range scattered light from the full solar disk. This background level is \pry{11.5}~DN~s$^{-1}$~pix$^{-1}$, derived from the $D_{\rm ann}=0$ value of the linear fit. The average $D_{\rm V}$ value from the PSF-deconvolved images for the on-disk data points (\textit{blue line} in Figure~\ref{fig.aia-int}a) is 14.5~DN~s$^{-1}$~pix$^{-1}$, giving confidence that the long-range scattered light component can be  estimated from the linear fit to the $D_{\rm V}$--$D_{\rm ann}$ relation.

We can express the long-range scattered light at Venus as a fraction, $f$, of the full disk 193~\AA\ intensity, $D_{\rm fd}$, by first noting that the latter is relatively stable over time-scales of around a day. For example, extracting a 5-min cadence 193~\AA\ light curve for 2012 June 4 (the day prior to the transit) following the procedure described in Sect.~7.7.4 of \citet{sdo_guide} shows a standard deviation of only 4.4\%. From full disk images at 21:04~UT on June 5 and 05:58~UT on June 6, we obtain average intensities of 285 and 293~DN~s$^{-1}$~pix$^{-1}$ for the spatial region extending to 1.05~R$_\odot$ from Sun center. Taking the average of these two values gives $D_{\rm fd}=289$~DN~s$^{-1}$~pix$^{-1}$. We then have $f=11.5/289=0.0398$.

We therefore use the linear fit derived from  the Venus transit observations to suggest a general formula for the 193~\AA\ scattered light that can be applied to \textit{any} on-disk observation:
\begin{equation}\label{eq.aia}
    D_{\rm scatt} = {D_{\rm ann} \over 9.4} + {D_{\rm fd} \over 25.0}
\end{equation}
The factor 9.4 is  the reciprocal of the line gradient from Figure~\ref{fig.aia-int}, and the factor 25.0 is the reciprocal of $f$. We will adopt a similar expression for the EIS scattered light in Sect.~\ref{sect.eis-anal}.

\begin{figure}[t]
    \centering
    \includegraphics[width=0.6\textwidth]{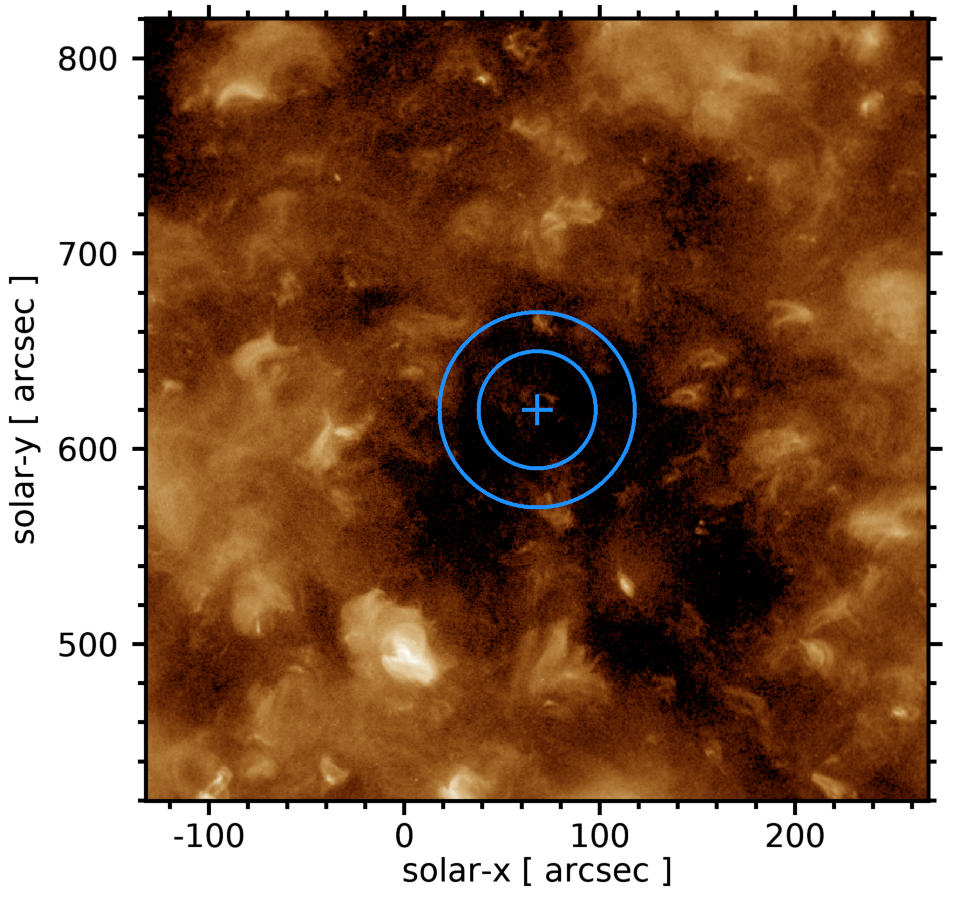}
\caption{An AIA 193~\AA\ image from 04:00:06~UT on 2016 November 23, shown with a logarithmic intensity scaling. The \textit{cross} denotes the location used as a check on the scattered light formula, the \textit{circles} show the annular region used to estimate the short-range scattered light component.}
\label{fig.aia-reg}
\end{figure}

As an example of the application of the formula, we choose the AIA 193~\AA\ image obtained at 04:00:06~UT on 2016 November 23 (Figure~\ref{fig.aia-reg}). A small coronal hole location at position (+68,+620) has an intensity of 12.3~DN~s$^{-1}$~pix$^{-1}$ (averaged over a 5\arcsec\ $\times$ 5\arcsec\ block); $D_{\rm ann}$ is measured as 12.6~DN~s$^{-1}$~pix$^{-1}$ and $D_{\rm fd}$=122.4~DN~s$^{-1}$~pix$^{-1}$. Thus Equation~\ref{eq.aia} gives the short-range scattered light component as 1.3~DN~s$^{-1}$~pix$^{-1}$ and the long-range component as 4.9~DN~s$^{-1}$~pix$^{-1}$, with the total scattered light component  50\%\ of the measured intensity. For comparison, the deconvolved image yields an average intensity in the same location of 11.8~DN~s$^{-1}$~pix$^{-1}$, implying a short-range scattered light component of 0.5~DN~s$^{-1}$~pix$^{-1}$ (since our interpretation of the data in Figure~\ref{fig.aia-int} is that the deconvolution only removes the short-range scattered light).

%-------------------------------
\section{EIS observations of the transit}\label{sect.eis}

\pry{EIS is one of three science instruments on board the \hinode\ spacecraft, which was launched in 2006. It is an imaging slit spectrometer that offers a choice of four slits with different widths. Two slits have narrow widths that translate to angular widths of 1\as\ and 2\as\ on the Sun and are used for emission line spectroscopy. Two wide slits (or ``slots") have widths of 40\as\ and 266\as\ and result in images appearing on the detector at the location of each emission line. For strong lines, the 40\as\ slit is narrow enough to yield relatively clean images not affected by overlap with neighboring lines. Further details are given in \citet{2022arXiv220314161Y}. EIS obtains spectra in the two wavelength bands 170--212~\AA\ and 246--292~\AA, referred to as short-wavelength (SW) and long-wavelength (LW), respectively. The spatial resolution is 3--4\as\ \citep{2022arXiv220314161Y}, and spatial coverage along the slit direction is 512\as. A scanning mechanism enables images up to 800\as\ wide to be built up through rastering.
}

\hinode\  observations of the Venus transit were organized through \href{http://www.isas.jaxa.jp/home/solar/hinode_op/hop.php?hop=0209}{\hinode\  Operation Plan No.~209}, led by T.~Shimizu and A.~Sterling, and the EIS Chief Observer was K.~Aoki. 
The \hinode\  pointing system is not suitable for tracking Venus during the transit so a set of six fixed pointings was performed, with Venus drifting through the fields of view of the three \hinode\  instruments. The pointing changes were performed at a frequency of 98.5~min, corresponding to the orbital period of \hinode.  The observations \pry{occurred} during the eclipse season, so the available observing time per orbit for EIS was about 65~min.

\begin{deluxetable}{clcccc}[t]
\tablecaption{EIS observations for the Venus transit.\label{tbl.eis-slot}}
\tablehead{ 
Pointing No. & Study name & Start time & End time & Position & No. rasters}
\startdata
1  & SI\_Venus\_slot\_v1  & 21:05 & 21:07 & [-940,+559] & 1 \\
   & SI\_Venus\_slit      & 21:10 & 21:52 & [-1003,+559] & 8 \\
2  & SI\_Venus\_slot\_v1  & 22:48 & 23:33 & [-568,+559] & 20 \\
3  & SI\_Venus\_slot\_v1  & 00:07 & 01:14 & [-212,+559] & 30 \\
4  & SI\_Venus\_slot\_v1  & 01:45 & 02:52 & [+179,+559] & 30 \\
5  & SI\_Venus\_slot\_v2  & 03:23 & 03:43 & [+498,+559] & 6 \\
   & SI\_Venus\_slot\_v2  & 03:50 & 04:10 & [+598,+559] & 6 \\
6  & SI\_Venus\_slot\_v2  & 05:01 & 05:21 & [+895,+560] & 6 \\
   & SI\_Venus\_slit\_v2  & 05:30 & 05:49 & [+1019,+560] & 10 \\
\enddata
\end{deluxetable}

Table~\ref{tbl.eis-slot} lists the sequence of
EIS observations obtained for the Venus transit, which began at 21:05 and completed by 05:49 thus within the period studied with AIA in the previous section. The six \hinode\  pointings are indicated, along with the EIS study name, observation start and end time, the position of the raster centers, and the number of raster repeats. The four studies were designed by Dr.~S.~Imada, and they used either  the 40\arcsec\ slot or the 2\arcsec\ slit (indicated in the titles of the studies). Only data from the slot studies are considered here.

The \textsf{SI\_Venus\_slot\_v1} study is a 6-step
raster with 20~s exposure times and a raster duration of
2~min and 10~s. The \textsf{SI\_Venus\_slot\_v2} study is a 2-step raster with
100~s exposure times and a raster duration of 3~min and
22~s. \pry{Both studies download the same 10 wavelength windows from the two EIS channels, one of which yields \ion{Fe}{xii} \lam195.12. The other windows are discussed in Section~\ref{sect.ext}.}
%Both studies included the \ion{Fe}{xii} \lam195.12 emission line, which is the only one studied in the present work.

\begin{figure}[t]
    \centering
    \includegraphics[width=0.5\textwidth]{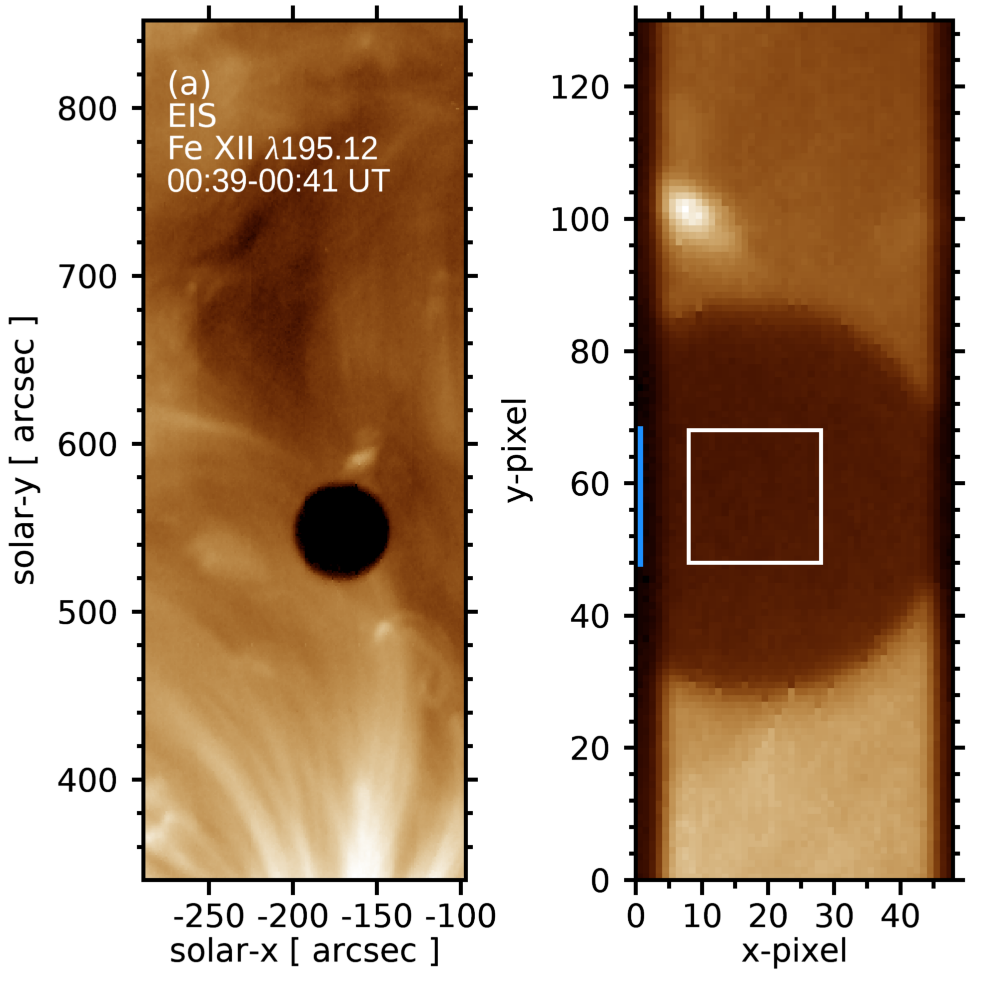}
\caption{An EIS \lam195.12 image derived from a raster obtained between 00:39 and 00:41~UT. A logarithmic intensity scaling has been applied. The shadow of Venus is visible at approximate position [$-170,+550$].}
\label{fig.eis-img}
\end{figure}

Figure~\ref{fig.eis-img}a shows a raster image from the \textsf{SI\_Venus\_slot\_v1} study beginning at 00:39~UT. It consists of six vertical strips of height 512\arcsec\ assembled from the six adjacent steps of the 40\arcsec\ slot, and  is built up from right to left. Figure~\ref{fig.eis-img}b shows the exposure from the third raster step as it appears on the EIS detector, with the $y$-range reduced to show the Venus shadow. Note that the image is reversed in the $x$-direction compared to Figure~\ref{fig.eis-img}a. The detector window used for \ion{Fe}{xii} \lam195.12 is 48 pixels wide, with 1 pixel corresponding to 1\arcsec.

\citet{2022arXiv220314161Y} performed a study of EIS slot data and found that the slot has a projected width on the detector of 41~pixels. Due to the line spread function of the instrument, the edges of the slot are blurred resulting in the slot image extending over 46~pixels, close to the width of the \lam195.12 wavelength window used for the transit observations.
\citet{2022arXiv220314161Y} found that intensities measured from the slot images in \ion{Fe}{xii} \lam195.12 are 14\%\ higher than those measured with the 1\as\ slit. They also highlighted the importance of subtracting a background level from the slot image to obtaining accurate intensities in quiet Sun and coronal hole regions. For EIS studies with a \lam195.12 wavelength window of 48 pixels in width they recommended using the leftmost and rightmost data columns in the window to represent the background.

%-------------------------------
\section{EIS data analysis}\label{sect.eis-anal}

\begin{figure}[t]
    \centering
    \includegraphics[width=\textwidth]{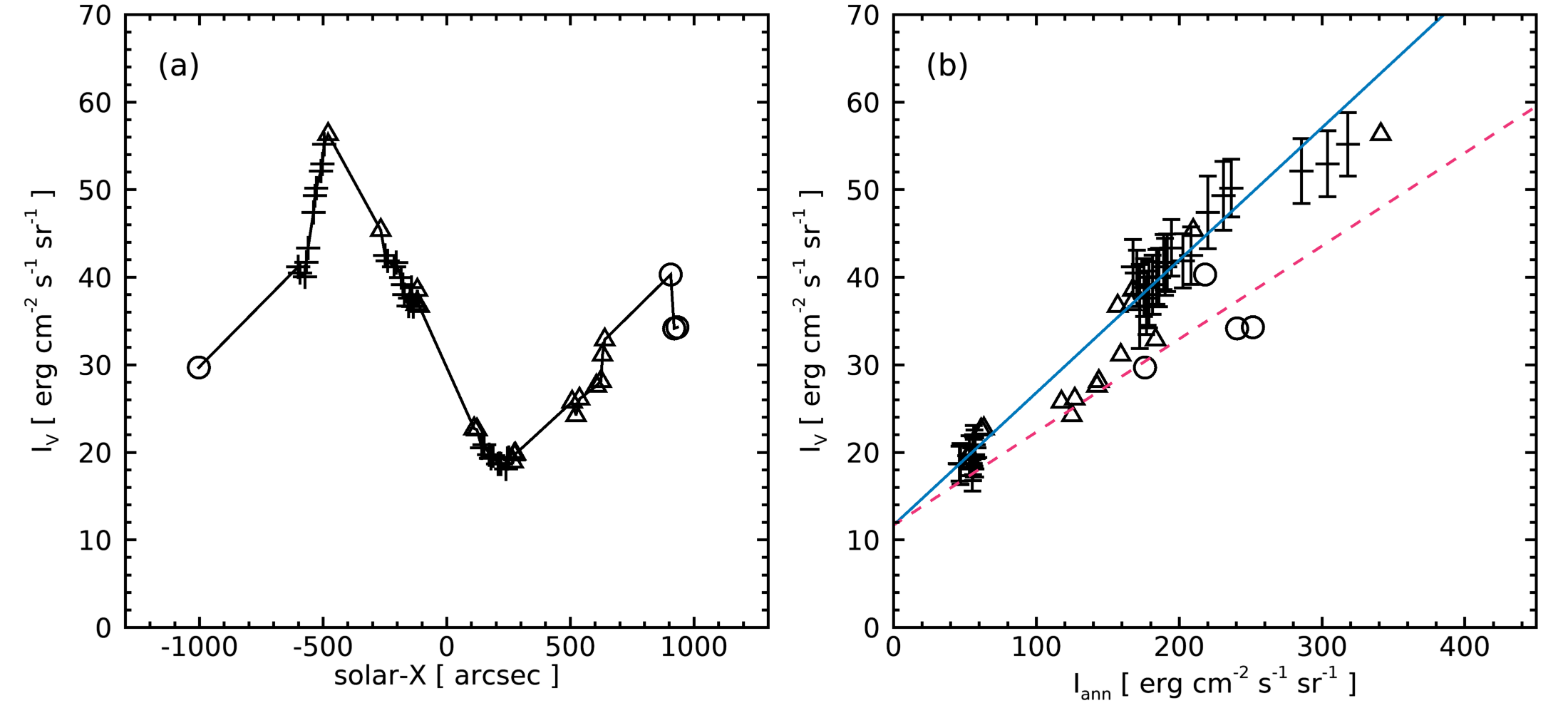}
\caption{Panel (a) shows the \ion{Fe}{xii} \lam195.12 Venus intensity, $I_{\rm V}$, measured from EIS as a function of the solar-$x$ position of Venus.  \textit{Circles} indicate positions above the limb, \textit{crosses} positions on the disk, and \textit{triangles} positions on the disk where a complete annular intensity, $I_{\rm ann}$, could not be measured. Panel (b) plots $I_{\rm V}$ vs.\  $I_{\rm ann}$. The \textit{solid line} is the fit to the points inside the limb, and the \textit{dashed line} is the fit from AIA data (Figure~\ref{fig.aia-int}), normalized to cross the EIS line at $I_{\rm ann}=0$.}
\label{fig.eis-int}
\end{figure}

In this section \ion{Fe}{xii} \lam195.12 intensities are measured from the EIS Venus data in order to derive a scattered light formula similar to that for AIA (Equation~\ref{eq.aia}). We first note that users have a choice in selecting an absolute radiometric calibration for their EIS data. The \href{http://solarb.mssl.ucl.ac.uk:8080/eiswiki/Wiki.jsp?page=EISCalibration}{EIS Wiki} recommends using either the \citet{2013A&A...555A..47D} or the \citet{2014ApJS..213...11W} calibrations, which give similar results. Both are updates to the original laboratory calibration of \citet{2006ApOpt..45.8689L}, which is the default option within the EIS software. For the present work, all intensities were computed with the \citet{2014ApJS..213...11W} calibration, and then reduced by 14\%\ as mentioned in the previous section.

Figure~\ref{fig.eis-int}a shows the Venus intensities, $I_{\rm V}$, derived from  \ion{Fe}{xii} \lam195.12 during the transit, and it can be compared with the AIA plot shown earlier (Figure~\ref{fig.aia-int}a).
$I_{\rm V}$ was derived with the following procedure which is implemented in the IDL routine \textsf{eis\_venus\_select}, available in the \textsf{papers/2022\_venus} repository. An image is constructed from the individual slot exposures, such as the one shown in Figure~\ref{fig.eis-img}a which is constructed from six exposures. By manually selecting the center of Venus in this image, the code identifies the exposure that contains most of the Venus shadow. A close-up image of the shadow in this exposure is then displayed to allow the center of Venus to be manually selected. If the center is too close to the edge of the slot image, then the raster is rejected. Otherwise, a block of $21\times 21$ pixels centered on the selected Venus position (see Figure~\ref{fig.eis-img}b) is extracted and averaged to yield an intensity, $I$. 
\pry{An uncertainty, $\sigma$, is obtained from the standard deviation of the intensities in the $21\times 21$ pixel block.}
From the same exposure, the section of the leftmost pixel column that corresponds to the $y$-positions of the $21\times 21$ Venus block (the vertical blue line in Figure~\ref{fig.eis-img}b) was extracted and averaged to yield $I_{\rm bg}$. 
\pry{The standard deviation of the intensities of the 21 pixels yields an uncertainty $\sigma_\mathrm{bg}$.}
We then have $I_\mathrm{V}=I-I_{\rm bg}$\pry{, and $\sigma_\mathrm{V}^2=\sigma^2+\sigma_\mathrm{bg}^2$}. This procedure was performed for all 99 slot rasters (Table~\ref{tbl.eis-slot}), and 45 datasets were rejected. Most of the latter were because the center of Venus was too close to the edge of the slot (as noted above). Additional datasets were rejected because of missing exposures or because the raster was begun during orbital twilight when the EUV spectrum is partially absorbed by the Earth's atmosphere. The positions of Venus for the selected datasets are shown on Figure~\ref{fig.aia-image} as blue crosses. The track is different from that of AIA due to SDO having a geosynchronous orbit while \hinode\ has a polar, Sun-synchronous, low-Earth orbit. The angle subtended by the two spacecraft at Venus  can be different by up to around 200\as. The results from \textsf{eis\_venus\_select} were output to the text file \textsf{results\_195.txt}, which is available in the \textsf{papers/2022\_venus} repository.

Figure~\ref{fig.eis-int}b plots $I_\mathrm{V}$ against the annulus intensity, $I_{\rm ann}$, analogous to Figure~\ref{fig.aia-int}b for the AIA data. $I_{\rm ann}$ is computed as part of the \textsf{eis\_venus\_select} procedure through a call to the routine \textsf{eis\_get\_annulus\_int}. The latter calls \textsf{eis\_slot\_map} to create an IDL map with the \textsf{/bg48} keyword set, which removes the background intensity using the \citet{2022arXiv220314161Y} prescription for 48-pixel windows. The pixels between two circles of radius 30\arcsec\ and 50\arcsec, centered on the user-selected Venus position, are identified and averaged to then yield $I_{\rm ann}$. For many of the rasters, the annulus extends beyond the slot raster image, either because the Venus shadow is close to the raster edge, or because of the narrow width of the \textsf{SI\_Venus\_slot\_v2} rasters. The software computes the number of pixels entering into the annulus intensity calculation and prints the ratio ($R_\mathrm{ann}$) relative to the maximum number of pixels (5027) to the results file. Where $R_\mathrm{ann}<0.75$, the points in Figure~\ref{fig.eis-int} are plotted as triangles. All of the points obtained above the limb have $R_\mathrm{ann} <0.75$, and they are indicated with circles in Figure~\ref{fig.eis-int}.

Figure~\ref{fig.eis-int}a shows significant differences from the equivalent AIA plot (Figure~\ref{fig.aia-int}a). In particular, EIS did not observe Venus transiting the limb, with all observed Venus locations being at least 100\as\ from the limb. The peak seen at $x=-500$ in Figure~\ref{fig.eis-int}a arises from Venus passing close to the bright active region loops on the east side of the active region complex and a plume-like structure at around $y=600$\as\ (Figure~\ref{fig.aia-image}). A similar peak is not seen for AIA as Venus tracked further to the north compared to the EIS data. The plot of $I_\mathrm{V}$ vs.\ $I_\mathrm{ann}$ shows a larger spread of values compared to the equivalent AIA plot (Figure~\ref{fig.aia-int}a). However, the points for which $R_\mathrm{ann}\ge 0.75$ do show a clear linear trend, giving confidence that the linear relation found for AIA also applies to EIS. A linear fit to these points is over-plotted on Figure~\ref{fig.eis-img}b as a \textit{blue line}. \pry{The gradient of this line is $0.151\pm 0.006$.  Extrapolating the fit to $I_{\rm ann}=0$ gives a Venus intensity of $I_{\mathrm{V}_0}=11.7\pm 0.9$~\ecss}. This is the long-range component of the scattered light within the Venus shadow. 

Also shown in Figure~\ref{fig.eis-img}b is the linear fit from the AIA data, scaled to intersect the EIS linear fit at $I_{\rm ann}=0$. The EIS fit has a steeper gradient, suggesting that the local scattered light is a larger factor for EIS, which may be due to the different optical configurations of the instruments.

In analogy with the AIA analysis, we can write the EIS scattered light contribution during the Venus transit as a combination of the short-range component and the long-range component:
\begin{equation}\label{eq.eis1}
    I_{\rm V} = { I_{\rm ann} \over \alpha } + {I_{\rm fd}\over \beta}.
\end{equation}
$\alpha$ is simply the inverse of the gradient of the blue line in Figure~\ref{fig.eis-img}b, and thus $\alpha=6.6$. To determine $\beta$ we need an estimate of the full-disk \lam195.12 intensity during the transit. Full-disk measurements are not available from EIS, but it is possible to make use of the AIA 193~\AA\ full-disk intensity to obtain an estimate. The procedure is as follows.

Three slot rasters beginning at 23:16, 00:30 and 02:05~UT during the transit were selected. For each a full-disk AIA 193~\AA\ synoptic image was downloaded, with a time close to the EIS observation. (Since AIA interleaved partial frame images with full-disk images during the transit, the nearest-in-time AIA image was not necessarily a full-disk image.) Sub-images from the AIA images were extracted to match the EIS raster fields-of-view, and they were manually co-aligned by matching bright points in the images.  Co-spatial blocks of size 150\arcsec\ $\times$ 150\arcsec\ to the north of Venus were extracted and the AIA and EIS intensities were averaged over these blocks to give intensities $D_\mathrm{block}$ and $I_\mathrm{block}$, respectively.  The EIS intensities were derived following the procedure for the annulus intensity described earlier. The AIA images were used to obtain the full-disk intensities, $D_{\rm fd}$, averaged out to 1.05~R$_\odot$. The EIS full-disk \lam195.12 intensity \pry{[$I_\mathrm{fd}$]} is then approximated by $D_{\rm fd}I_\mathrm{block}/D_\mathrm{block}$, \pry{and $\beta=I_\mathrm{fd}/I_{\mathrm{V}_0}$}. The values of these parameters for the three datasets are given in Table~\ref{tbl.beta}. All of these numbers are generated with the IDL routine \textsf{eis\_full\_disk\_scale}, available in the \textsf{papers/2022\_venus}  repository. The average value of $\beta$ is \pry{34}, and this is used in the following section.

Appendix~\ref{app.hop130} compares the $I_\mathrm{fd}$ value derived using this method with the true  value obtained from a full-disk EIS scan performed on 2012 May 30, six days prior to the transit. The true $I_\mathrm{fd}$ value was found to be 13\%\ lower than the derived value, and demonstrates that our method of inferring the full-disk \lam195.12 intensity is reasonably accurate.

\begin{deluxetable}{llllll}
\tablecaption{AIA ($D$) and EIS ($I$) intensity measurements used to derive the parameter $\beta$.\label{tbl.beta}}
\tablehead{\colhead{Time} &
    \colhead{$D_\mathrm{block}$\tablenotemark{a}} &
    \colhead{$D_{\rm fd}$\tablenotemark{a}} &
    \colhead{$I_\mathrm{block}$\tablenotemark{b}} &
    \colhead{$I_{\rm fd}$\tablenotemark{b}} &
    \colhead{$\beta$}
}
\startdata
23:16 & 222 & 284 & 312 & 399 &  34.1 \\
00;31 &  85 & 287 & 119 & 403 &  34.4 \\
02:06 &  78 & 291 & 101 & 376 &  32.1 \\
\enddata 
\tablenotetext{a}{Units: DN~s$^{-1}$~pix$^{-1}$.}
\tablenotetext{b}{Units: \ecss.}
\end{deluxetable}

%-------------------------------
\section{Prescription for estimating scattered light in EIS data}\label{sect.pres}

The previous section gave the expression for the \ion{Fe}{xii} \lam195.12 scattered light intensity within the Venus shadow during the transit in terms of short-range and long-range components. The former is proportional to the local annulus intensity, $I_\mathrm{ann}$, and the latter is proportional to the full-disk \lam195.12 intensity, $I_\mathrm{fd}$. We now assume this expression applies to any on-disk EIS observation. If an intensity $I$ is measured from an EIS raster observation, either with the narrow slits or the slots, there is a component due to scattered light that  is given by
\begin{equation}\label{eq.eis2}
    I_{\rm scatt} = { I_{\rm ann} \over 6.6 } + {I_{\rm fd}\over 34}.
\end{equation}
where $I_{\rm ann}$ and $I_{\rm fd}$ are measured co-temporally with $I$. The former can usually be measured directly from the same EIS raster if the field-of-view is large enough. The latter must be derived from an AIA 193~\AA\ full-disk image, by cross-calibrating the intensity in a region within the EIS raster to the same region in the AIA image, as described in the previous section.

Here we summarize the procedure to estimate the scattered light component in an EIS raster.
\begin{enumerate}
    \item Measure $I_{\rm ann}$ for the location of interest from an EIS map. The IDL routine \textsf{eis\_annulus\_int} in the \textsf{aia-eis-venus} repository is provided for this purpose.
    \item Derive $I_{\rm fd}$ by comparing a region observed by EIS with that observed in the AIA 193~\AA\ channel. The IDL routine \textsf{eis\_aia\_int\_compare} in the \textsf{aia-eis-venus} repository is provided for this purpose, and generally a quiet Sun region with fairly uniform emission should be selected.
    \item Apply Equation~\ref{eq.eis2} to determine $I_{\rm scatt}$, and compare it with the intensity measured at the location of interest.
\end{enumerate}

How accurate will the scattered light estimates be? 
Some sources of uncertainty can be quantified. \pry{For example, the linear fit to the Venus intensities (Figure~\ref{fig.eis-int}b) yields uncertainties of 4\%\ and 8\%\ for the short- and long-range scattered light intensities.}
%has residuals with a standard deviation of 5\%. 
The AIA 193~\AA\ full-disk intensity may not be an accurate proxy of the \lam195.12 full-disk intensity, which would lead to uncertainties in both the parameter $\beta$ (Equation~\ref{eq.eis1}) and $I_{\rm fd}$ (Equation~\ref{eq.eis2}). Appendix~\ref{app.hop130} performs a check on the AIA--EIS calibration method using a full-disk \lam195.12 measurement from 2012 May 30, and finds that the method under-estimates the \lam195.12 intensity by 13\%. This discrepancy may vary with solar conditions as the spectral content of the AIA 193~\AA\ varies with changing solar activity (e.g., greater or lesser contributions from species cooler or hotter than \ion{Fe}{xii} to the channel). This is likely to be small given that the entire corona emits strongly in \ion{Fe}{xii}, however. The following section demonstrates that for two coronal hole rasters the scattered light formula predicts an intensity larger than the measured intensity. The worst case suggests an uncertainty of at least 16\%. Overall, we suggest the scattered light estimate is accurate to around \pry{25}\%.

\pry{One scenario where the formula will underestimate the scattered light is when there is a bright active region close to the point of interest, but outside the 50\as\ radius of the annulus. This is explored in Appendix~\ref{app.ar-model}, where it is found that a bright active region enhances the short-range scattered light by around 50\%\ if it is located at 100\as\ from the point of interest. Features that may be impacted are the low-intensity patches at the edges of active regions that demonstrate outflowing plasma \citep{2007Sci...318.1585S}. The case considered in Appendix~\ref{app.ar-model} was deliberately chosen to be an extreme example, given the brightness of the active region. Applying a deconvolution algorithm to an AIA 193~\AA\ co-temporal with the EIS observation of interest, such as done in Appendix~\ref{app.ar-model}, would give an indication of the effect of the AR on the EIS data.
}

If the EIS field-of-view is too small to enable the annulus intensity to be measured, or the field-of-view is compromised by missing data, then the suggested solution is to use AIA 193~\AA\ as a proxy. \pry{The EIS location within the AIA image should be identified and  the annulus intensity from the co-temporal AIA 193~\AA\ image obtained with the IDL routine \textsf{aia\_annulus\_int}}. The EIS/AIA quiet Sun calibration factor from Step (2) can then be used to convert the AIA annulus intensity to an EIS intensity. For coronal holes this will likely be an over-estimate of the EIS intensity as the 193~\AA\ channel has contamination from cooler species such as \ion{Fe}{vii} and \ion{Fe}{viii} that is more pronounced in coronal holes.

If the raster does not include a suitable patch of quiet Sun for Step (2) then another raster can be used. The stability of the AIA 193~\AA\ full-disk emission with time means that an observation within $\pm$~1~day of the raster of interest should give good results.

AIA has given almost continuous, high-cadence, full-disk coverage of the Sun since 2010~May. Prior to this time an EIT 195~\AA\ image can be substituted in order to provide the calibration necessary to yield the EIS full-disk intensity. The routine \textsf{eis\_aia\_int\_compare} automatically searches for an EIT image if an AIA image is not available.

One caveat of the prescription is that it does not account for the local scattered light coming from sources inside the inner radius of the annulus.  For this reason, we recommend that our prescription only be applied if the intensity within the inner boundary is relatively uniform. Otherwise the estimated scattered light will only be a lower limit.

\begin{figure}[t]
    \centering
    \includegraphics[width=0.5\textwidth]{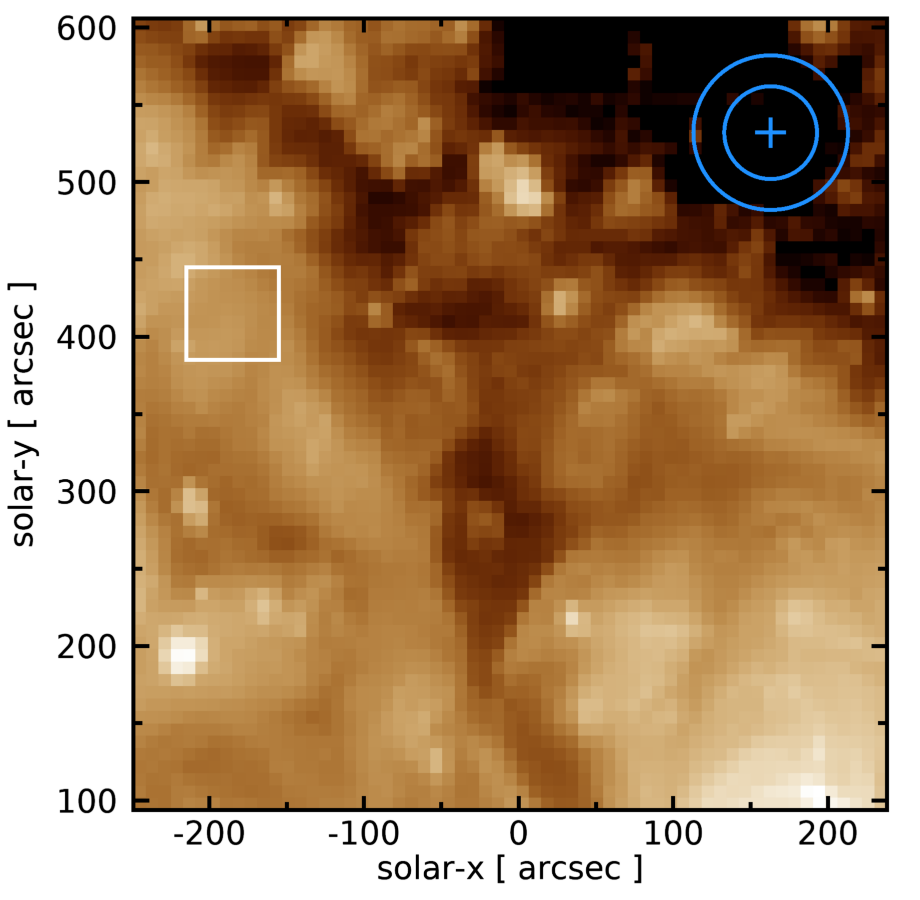}
\caption{A raster image from EIS obtained between 03:08 and 05:15~UT on 2016 November 23 and showing the intensity of \ion{Fe}{xii} \lam195.12 on a logarithmic scale. The dark region to the top-right is part of a coronal hole, and the \textit{circles} indicate the annular region used for obtaining the local scattered light component. The \textit{white box} indicates the quiet Sun region used for normalizing the \lam195.12 intensity with the AIA 193~\AA\ channel.}
\label{fig.reg}
\end{figure}

As an example of applying the formula, we consider an EIS raster that began at 03:08~UT on 2016 November 23. This is a narrow slit raster obtained with the study DHB\_006\_v2, and the raster image from \ion{Fe}{xii} \lam195.12 is shown in Figure~\ref{fig.reg}. The north-west corner of the raster contains a section of a coronal hole. Due to the low signal in this region, $2\times 8$ spatial binning was applied to the data. The \lam195.12 intensity is 5.7~\ecss\ at  position $(+163,+532)$ within  the coronal hole, indicated with a \textit{blue cross} on Figure~\ref{fig.reg}. The annulus intensity at this location (shown in Figure~\ref{fig.reg}) is measured as 7.5~\ecss, giving a short range scattered light component of 1.1~\ecss. To obtain the long-range scattered light component we first selected a quiet Sun region of 60\as\ $\times$ 60\as\ centered at $(-185,+415)$---also shown in Figure~\ref{fig.reg}---and calibrated it against an AIA 193~\AA\ image at 04:56~UT to yield a full-disk \lam195.12 intensity estimate of 187~\ecss. From Equation~\ref{eq.eis2}, the long-range component is then \pry{5.5}~\ecss. The estimated scattered light intensity is thus larger than the measured intensity in the coronal hole by \pry{16}\%, and so we infer that the \lam195.12 intensity is entirely due to scattered light at this position.

%----------------------------
\section{Scattered light in coronal holes}\label{sect.ch}

Using the browse products at the \href{https://eismapper.pyoung.org}{\textit{EIS Mapper}} website \citep{young_peter_r_2022_6574455}, a number of coronal hole observations between 2011 and 2018 were identified and processed to estimate the percentage of scattered light at coronal hole locations. In each case the rasters had sufficient spatial coverage to allow the annulus intensity to be measured. The EIS-AIA calibration was performed by selecting quiet Sun regions of fairly uniform intensity in the same rasters, with the exceptions of the 2010 and 2013 datasets for which it was necessary to use another slit raster obtained on the same day. Details are given in the supplementary material provided in the \textsf{GitHub} \href{https://github.com/pryoung/papers/tree/main/2022_venus}{\textsf{pryoung/papers/2022\_venus}} repository.

\begin{deluxetable}{cclcDDc>{\bfseries}D>{\bfseries}D>{\bfseries}c}[t]
\tablecaption{Intensities for seven EIS coronal hole observations.\label{tbl.ch}}
\tablehead{ 
    \colhead{Date} &  
    \colhead{Time} & 
    \colhead{Study} &  
    \colhead{Position} & 
    \twocolhead{$I_{\rm ch}$} & 
    \twocolhead{$I_{\rm ann}$} & 
    \colhead{$I_{\rm fd}$} & 
    \twocolhead{Short} & 
    \twocolhead{Long} & 
    \colhead{\%} 
}
\decimals
\startdata
17-May-2010 & 12:19 & \textsf{Atlas\_60} & $(-98,-476)$ & 8.3 & 9.7 & 188 
     & 1.5 & 5.5 & 84\%\\
12-Jan-2011 & 00:09 & \textsf{YKK\_EqCHab\_01W} & $(+119,+312)$ & 12.7 & 14.6 & 312 & 2.2 & 9.2 & 90\%\\
31-Jan-2013 & 06:12 & \textsf{YKK\_ARabund01}  & $(+52,+47)$ & 26.3 & 49.5 & 391 & 7.5 & 11.5 & 72\%\\
20-Jun-2015 & 11:18 & \textsf{GDZ\_PLUME1\_2\_300\_50s}  & $(-140,-240)$ & 27.1 & 31.7 & 354 & 4.8 & 10.4 & 56\% \\
23-Nov-2016 & 03:08 & \textsf{DHB\_006\_v2} & $(+163,+532)$ & 5.7 & 7.5 & 192 & 1.1 & 5.5 & 116\% \\
13-Jun-2017 & 04:55 & \textsf{DHB\_007} & $(-138,-124)$ & 7.8 & 10.6 & 198 &
     1.6 & 5.8 & 95\%\\
18-Aug-2018 & 23:21 & \textsf{HPW021\_VEL\_240x512v2} & $(-20,-445)$ & 
     6.8 & 18.5 & 208 & 2.8 & 4.6 & 109\% \\
\enddata
\end{deluxetable}

Table~\ref{tbl.ch} gives the measured intensities for each dataset, and the short and long-range scattered light components estimated from Equation~\ref{eq.eis2}. The final column gives the percentage contribution of scattered light to the measured coronal hole intensity ($I_\mathrm{ch}$). \pry{It can be seen that the scattered light component is dominant for all seven datasets, and makes a contribution of 90\%\ or more for four datasets.}
%It can be seen that for five of the seven datasets, the coronal hole intensity is entirely or almost entirely due to scattered light. Even for the remaining two datasets (2013 and 2015) the scattered light component is dominant. 
The long-range scattered light component is the most important in all cases. Thus, even for a coronal hole dataset in the heart of a large coronal hole where there are no nearby bright emission sources, there will always be a significant scattered light component to \ion{Fe}{xii} \lam195.12.

%----------------------------
\section{Extension to Other EIS Wavelengths}\label{sect.ext}

\pry{
The prescription described in Section~\ref{sect.pres} applies specifically to \ion{Fe}{xii} \lam195.12, which is found in the EIS SW channel. A similar formula to Equation~\ref{eq.eis2} would be expected to apply to other lines but the $\alpha$ and $\beta$ (Equation~\ref{eq.eis1}) parameters may be different. A particular concern is whether there is a wavelength dependence that may be significant for lines in the EIS LW channel, as the \ion{S}{x} and \ion{Si}{x} ions used for FIP bias measurements (Section~\ref{sec:intro}) have lines in the 258--265~\AA\ region of the EIS LW channel.
%In particular, ions formed at the same temperature as \ion{Fe}{xii}, such as the \ion{S}{x} and \ion{Si}{x} ions that are used for abundance diagnostics may have different scattering formulae due to their longer wavelengths. 
}

\pry{
Formulae for other lines can potentially be derived from the EIS Venus observations. Nine wavelength windows were used in addition to the one for \ion{Fe}{xii} \lam195.12. These were centered on the following lines: \ion{Fe}{xi} \lam180.40, \ion{O}{vi} \lam184.12, \ion{O}{v} \lam192.90, \ion{He}{ii} \lam256.32, \ion{Fe}{xvi} \lam262.98, \ion{Mg}{vi} \lam269.00, \ion{Fe}{xiv} \lam274.20, \ion{Si}{vii} \lam275.35 and \ion{O}{iv} \lam279.93. The \ion{O}{iv} and \ion{Mg}{vi} lines are too weak to be useful, while \ion{O}{v} and \ion{O}{vi} are affected by blends with strong nearby lines. 
}

\pry{
\ion{Fe}{xvi} has $\log\,(T_\mathrm{f}/\mathrm{K})= 6.45$, which means that it has negligible emission from quiet Sun and coronal holes. Thus any scattered light measured in these regions can only come from nearby active regions and there can be no long-range component comparable  to that for \lam195.12.
The remaining ions are \ion{He}{ii}, \ion{Fe}{xi} and \ion{Fe}{xiv} lines. Of these, the latter is of the most interest as the wavelength is furthest from 195.12~\AA\  and offers the opportunity of checking if the scattering formula shows some dependence on wavelength.
}

\begin{figure}
    \centering
    \includegraphics[width=\textwidth]{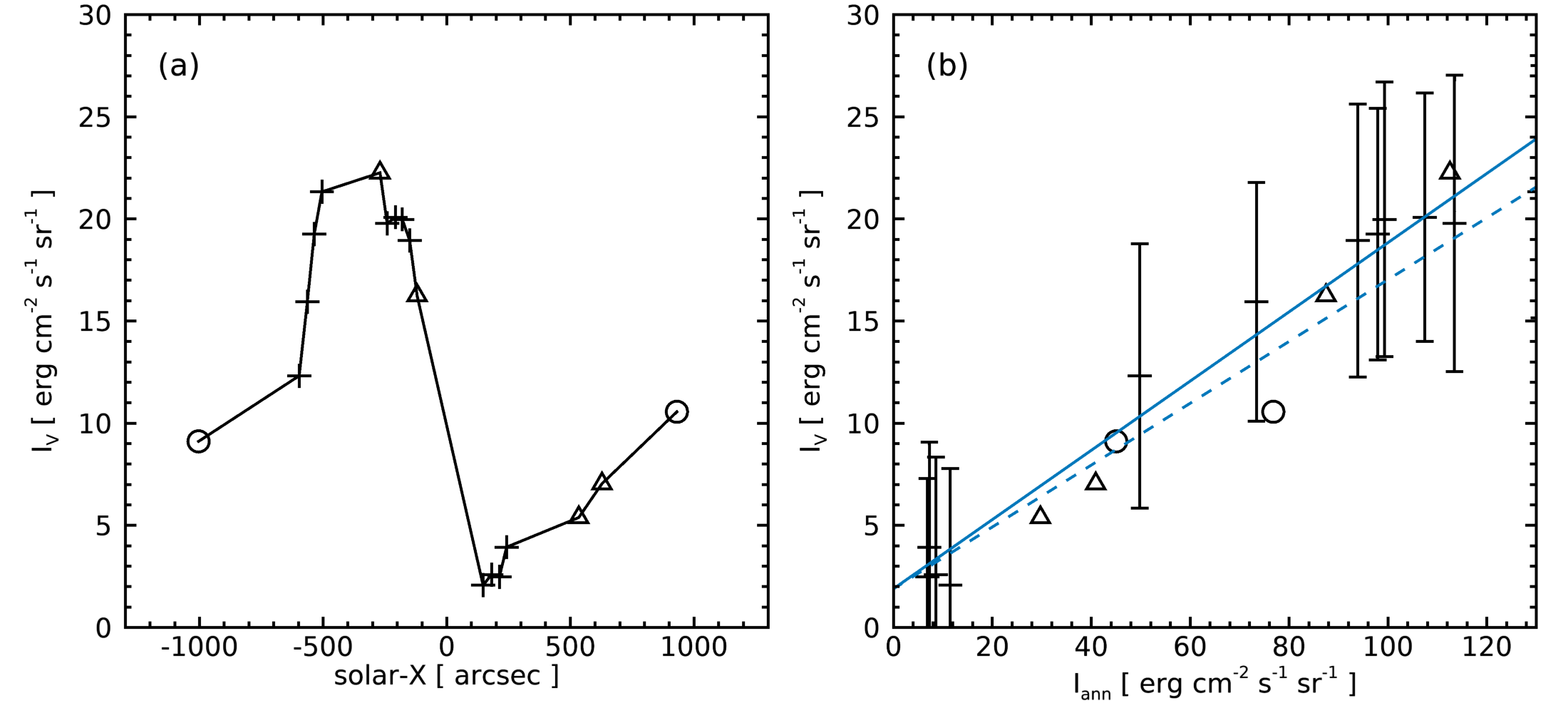}
    \caption{Plots of the Venus \ion{Fe}{xiv} \lam274.20 intensity during the eclipse. An explanation of the symbols is given in Figure~\ref{fig.eis-int}. The blue dashed line on Panel (b) gives the linear fit from \ion{Fe}{xii} \lam195.12, scaled to intersect the \lam274.20 fit at $I_\mathrm{ann}=0$.}
    \label{fig.274}
\end{figure}

\pry{
A reduced set of rasters compared to \lam195.12 were processed using the \textsf{eis\_venus\_select} routine and the results are presented in Figure~\ref{fig.274}, which can be compared to Figure~\ref{fig.eis-int} for \lam195.12. Immediately apparent are the large uncertainties for $I_\mathrm{V}$. The EIS effective area is about a factor five lower at 274.20~\AA\ compared to 195.12~\AA\ \citep{2014ApJS..213...11W}, and the \ion{Fe}{xiv} line generally has a lower intensity in quiet Sun and active region conditions \citep{2008ApJS..176..511B}. The linear fit to the data points gives a gradient of $0.169 \pm 0.037$ and $I_{\mathrm{V}_0}= 1.90\pm 2.77$. The latter is poorly constrained due to the very low signal near the coronal hole.
}

\pry{
The gradient of the linear fit is close to that found for \lam195.12 (shown graphically in Figure~\ref{fig.274}), which suggests that the behavior of the short-range scattered light does not vary much as a function of wavelength. No statement about the wavelength dependence of the long-range scattered light, which is partly determined by $I_{\mathrm{V}_0}$, can be made due to the large uncertainties. For reference, however, we note that the AIA 211~\AA\ channel can be used to estimate the \ion{Fe}{xiv} full-disk intensity as it is usually dominated by a strong \ion{Fe}{xiv} line at 211.32~\AA\ \citep{2010A&A...521A..21O}. The \lam274.20/\lam211.32 ratio is insensitive to plasma conditions, with a ratio around 0.5. Performing a scaling using the Venus transit data similar to that discussed for \lam195.12 in Section~\ref{sect.eis-anal} gives a full-disk \lam274.20 intensity of 196~\ecss.  
A comparison of an AIA 211~\AA\ image and an EIS full-disk scan (Appendix~\ref{app.hop130}) shows that the method of estimating the full-disk \lam274.20 intensity using the AIA 211~\AA\ channel over-estimates the true \lam274.0 intensity by only 14\%. If we assume the value of $\beta$ derived for \lam195.12 also applies to \lam274.20, then the long-range scattered light expected for \lam274.20 is $196/34= 5.8$~\ecss. This is outside the $1\sigma$ uncertainty for  $I_{\mathrm{V}_0}$ but, given the uncertainties in our method described in Section~\ref{sect.pres} we do not find evidence for a significantly different scattered light formula for \lam274.20.
}

\begin{figure}[t]
    \centering
    \includegraphics[width=0.6\textwidth]{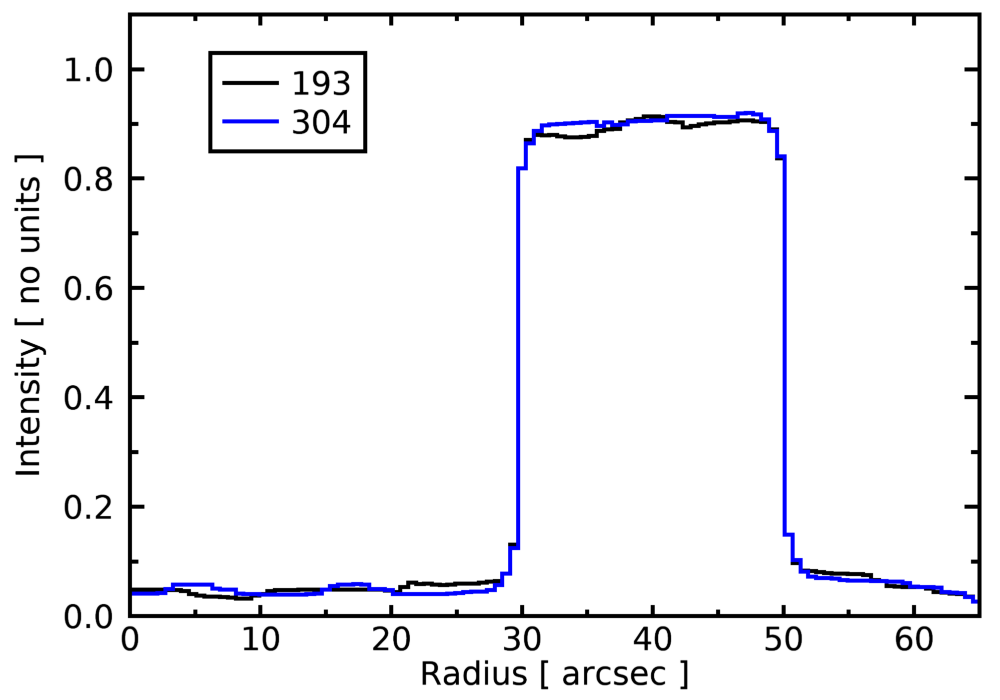}
    \caption{Radial slices through synthetic AIA 193~\AA\ and 304~\AA\ images derived by convolving annulus intensity distributions with the PSFs of \citet{aia-psf}.}
    \label{fig.193-304}
\end{figure}

\pry{
Another approach to investigating the dependence of scattered light on wavelength is to consider the \citet{aia-psf} prescription for scattered light in the AIA instrument. We consider an artificial AIA image that is zero everywhere, except for an annulus of inner radius 30\as\ and outer radius 50\as\ that has a uniform intensity of one. We then convolved this with the PSF functions for the 193 and 304~\AA\ channels, noting the mesh pitch is very similar for the two channels: 70.4 lines/inch and 70.2 lines/inch for 193~\AA\ and 304~\AA, respectively. Figure~\ref{fig.193-304} shows intensity cuts through the convolved images. Averaging the intensities over 5\as\ radius circles at the center of the annulus gives a 304~\AA\ intensity that is 10\%\ higher than that for 193~\AA. If we assume similar behavior for EIS, for which there is a single mesh for all wavelengths, then we may expect the scattered light arising from the mesh to be up to 10\%\ larger for the lines in the EIS LW channel compared to \lam195.12. This is within the 25\%\ uncertainty that we quoted for the \lam195.12 scattering formula. The mesh scattered light does not explain the long-range scattered light component, which may show a different behavior with wavelength, but this can not be explored with the current data.
}

\pry{
In summary, although we expect some wavelength variation to the degree of scattered light for different wavelengths observed by EIS, our investigation of the \ion{Fe}{xiv} \lam274.20 line and the 193~\AA\ and 304~\AA\ filters of AIA suggests that this variation is within the uncertainties for our formula derived from \ion{Fe}{xii} \lam195.12. The formula can thus be applied to other EIS lines except for the following exceptions. Firstly, there is a requirement that the full-disk intensity of an emission line can be estimated using images from AIA, or another full-disk imaging instrument. The EUV AIA 94~\AA, 131~\AA, 171~\AA, 193~\AA, 211~\AA, 304~\AA\ and 335~\AA\ channels are dominated by \ion{Fe}{x}, \ion{Fe}{viii} , \ion{Fe}{ix}, \ion{Fe}{xii}, \ion{Fe}{xiv}, \ion{He}{ii} and \ion{Fe}{xvi}, respectively, in non-flaring conditions \citep{2010A&A...521A..21O}. So only full-disk intensities from emission lines of these ions, or ions formed at the same temperatures as these ions can be estimated. A second exception is 
for lines with $\log\,(T_\mathrm{f}/\mathrm{K})\ge 6.3$, for which the long-range scattered light component will be inaccurate due to very low emission from the quiet sun. 
}

%\ion{Fe}{xiv} \lam274.20 is significantly weaker than \ion{Fe}{xii} \lam195.12 due to a factor five difference in effective area \citep{2014ApJS..213...11W}. However, the Venus shadow is clearly visible in most of the EIS slot rasters, except for the set beginning at 01:45~UT (Table~\ref{tbl.eis-slot}) when the shadow passed close to the low-latitude coronal hole

%For \ion{Fe}{xii} \lam195.12, the long-range scattered light is expressed as a fraction of the full-disk intensity, derived using the AIA 193~\AA\ filter. The AIA 211~\AA\ filter is centered on the strongest \ion{Fe}{xiv} line at 211.32~\AA\ and so potentially a similar approach can be applied to EIS \ion{Fe}{xiv} \lam274.20. The \lam274.20/\lam211.32 ratio is insensitive to plasma conditions, varying between 0.50 and 0.55 over typical coronal densities.

%----------------------------
\section{Summary and Discussion}\label{sect.summary}

An empirical formula for estimating the scattered light in an EIS raster for the \ion{Fe}{xii} 195.12~\AA\ emission line has been presented (Equation~\ref{eq.eis2}), based on data obtained during the Venus transit of 2012 June 5--6. \pry{Evidence from  EIS \ion{Fe}{xiv} \lam274.20 and a consideration of the AIA 193~\AA\ and 304~\AA\ channels (Section~\ref{sect.ext}) suggests that the \lam195.12 scattering formula can be applied to other emission lines in the EIS wavelength ranges.} A prescription is provided (Section~\ref{sect.pres}) to enable users to assess the effect of scattered light for any EIS observation that makes use of full-disk images obtained by either AIA or EIT. 

The intended application of the formula is an estimate of scattered light in on-disk coronal hole observations, and the results of Table~\ref{tbl.ch} show that scattered light dominates for low  \ion{Fe}{xii} intensities in coronal holes. If the coronal hole intensity is brighter, such as for the 2013 and 2015 examples, then the true coronal hole intensity can be significant. Bright structures within coronal holes such as plumes will be less affected, as should the coronal hole boundary regions that have intensities intermediate between quiet Sun and coronal hole.

% Although the formula and results presented here specifically apply to the \ion{Fe}{xii} line, the results are applicable to other species formed at the same temperature. Of particular interest are the ions \ion{S}{x} and \ion{Si}{x} that are commonly used as a diagnostic of the FIP bias. 

\pry{Based on our conclusion that the scattered light formula should apply to other lines in the EIS wavebands, the coronal hole results for \ion{Fe}{xii} extend to the \ion{S}{x} and \ion{Si}{x} ions, which are formed at the same temperature.}
\citet{2011ApJ...727L..13B} used the \ion{S}{x} and \ion{Si}{x} lines to measure the FIP bias in eight polar coronal hole observations finding, on average, no FIP bias. The time period of the datasets was not given, but was likely to be 2007--2008. \citet{2013ApJ...778...69B} studied an active region within a low-latitude coronal hole observed in 2007 October and also found no evidence of a FIP bias in the coronal hole. The results from Table~\ref{tbl.ch} suggest that the coronal hole intensities of  \ion{S}{x} and \ion{Si}{x} largely come from long-range scattered light. This component would be expected to show the average FIP bias of the entire solar disk. The true coronal hole intensity (if any) and the short-range scattered light component will have the actual coronal hole FIP bias. \citet{2015NatCo...6.5947B} derived a FIP bias map of the entire solar disk from an observation in 2013 January. Values ranged from 1.5 to 2.5 over most of the disk, with the lower values generally corresponding to lower intensity regions. Five low intensity regions that are probably coronal holes are seen on the Sun in the intensity maps. These regions do not show a FIP bias of 1, and they can barely be distinguished from quiet Sun in the FIP bias map. We consider this to be consistent with our conclusion  that the coronal hole intensities have a significant component of scattered light. The earlier measurements of coronal hole FIP biases close to one may be due to the global corona having a lower FIP bias during the solar minimum period of 2007--2008, although this is speculation. 

The 2015 observation from Table~\ref{tbl.ch} is an example of a relatively high intensity within the coronal hole, with the long-range scattered light only contributing 33\%\ to the measured intensity. A low-latitude coronal hole was observed and it is not clear if such holes generally have higher intensities or if coronal holes are generally brighter around solar maximum. The 2013 coronal hole also had a high intensity and was another low-latitude coronal hole observation.

An additional consequence of the scattered light contribution to \ion{Fe}{xii} \lam195.12 in coronal holes is that  Doppler shifts may not reflect the true (if any) Doppler shifts in the coronal hole. \citet{2010ApJ...709L..88T} presented a Doppler map in \ion{Fe}{xii} \lam195.12 of the north coronal hole and blue-shifts of around 20~\kms\ are clearly seen around the coronal hole boundary. A follow-up paper \citep{2011ApJ...736..130T} clarified that these blueshifts are mostly due to quiet Sun plumes along the line-of-sight to the coronal hole. Our independent check of the darkest parts of the coronal hole, close to the limb, show no significant blueshifts, and this can be seen in the authors' Figure~1 around coordinate (130,300). Our interpretation is that these dark areas are dominated by scattered light from the full disk, and so show no blueshift.  \citet{2014ApJ...794..109F} measured Doppler shifts in plumes and compared them with nearby coronal hole and quiet Sun regions. For \ion{Fe}{xii} they found no significant difference between quiet Sun and coronal hole velocities. This result also supports our suggestion that coronal holes are dominated by scattered light and so \ion{Fe}{xii} can not be used to measure an outflow velocity (if it exists) in coronal holes. \citet{2018ApJ...856...28W} also noted that centroid maps in \ion{Fe}{xii} \lam195.12 for two low-latitude coronal holes do not show evidence for Doppler shifts, consistent with their suggestion that scattered light is important.

%Coronal holes are the main source of the solar wind, and remote sensing FIP bias measurements are potentially a means of identifying connections between the corona and solar wind, since the bias is set in the chromosphere 
%is considered an essential measurement for identifying a connection between in situ solar wind measurements and the source region in the corona (**ref). 
% The empirical formula will be important for assessing if the intensities in a given source region, if observed by EIS, are affected by scattered light.

Finally, we highlight that, as part of the present work, an empirical formula for scattered light in the AIA 193~\AA\ channel was also derived (Equation~\ref{eq.aia}) and this may be useful for scientists interested in assessing the effect of long-range scattered light in their data.

\begin{acknowledgements}
The authors acknowledge support from the GSFC Internal Scientist Funding Model competitive work package program ``Connecting the corona to solar wind structure and magnetospheric impact using modeling and remote and in situ observations". P.~Young also acknowledges funding from the \hinode\  project. The authors thank I.~Ugarte-Urra for providing measurements from HOP~130 and for giving valuable comments on an early version of the manuscript. The anonymous referee is also thanked for insightful comments that improved the article.
\end{acknowledgements}

\facilities{Hinode (EIS), SDO (AIA)}

\bibliography{ms}{}
\bibliographystyle{aasjournal}

%==================================================
\appendix

%------------------------------
\section{Comparison of full-disk intensities with HOP 130}\label{app.hop130}

HOP 130 (PI: I.~Ugarte-Urra) is run every three weeks and combines large-format 40\as\ slot rasters 
%(EIS study \textsf{full\_sun\_slot\_scan\_1}) 
with  multiple spacecraft pointings to obtain full coverage of the solar disk. \pry{These data enable a true estimate of the average full-disk solar intensity in a specific emission line that can be compared with the intensity inferred via the method described in Section~\ref{sect.eis-anal}.}

The nearest-in-time run of HOP 130 to the Venus transit was performed on 2012 May 30, six days prior to the transit. Dr.~Ugarte-Urra processed this data-set to yield a \ion{Fe}{xii} \lam195.12 full-disk intensity of 463~\ecss, which has been averaged over the disk out to 1.05~R$_\odot$. The \citet{2014ApJS..213...11W} calibration was used. The EIS study used 40-pixel wavelength windows and \citet{2022arXiv220314161Y} recommend dividing by a factor 1.27 to yield intensities that are consistent with the EIS narrow slits. This then gives a final intensity of 365~\ecss.

On the same day the narrow slit raster \textsf{HPW021\_VEL\_240x512v1} was run at 00:43~UT. A Gaussian fit was performed to \ion{Fe}{xii} \lam195.12 at each pixel of this raster, and the intensity was averaged in a 50\as\ $\times$ 50\as\ region centered at heliocentric coordinates ($-199,+12$) in a quiet Sun region. The routine \textsf{eis\_aia\_int\_compare} was then applied to yield an estimate of the full-disk \lam195.12 intensity of 412~\ecss, 13\%\ higher than the intensity obtained from the HOP 130 dataset. This demonstrates that the quiet Sun calibration method gives reasonable estimates of the full-disk \lam195.12 intensity.

\pry{
In Section~\ref{sect.ext} the EIS \ion{Fe}{xiv} \lam274.20 line from the transit is analyzed, and a similar approach of estimating the full-disk intensity of this line by using an AIA image is applied. The HOP~130 study run on 2012 May 30 did not include \lam274.20, but an updated version was run on 2013 April 11 that did include this line. The resulting full-disk intensity was 176~\ecss\ using the \citet{2014ApJS..213...11W} calibration. Assuming the 1.27 empirical correction factor also applies to \lam274.20, the corrected intensity is 139~\ecss.
}

\pry{
On the following day, the study \textsf{HH\_Flare+AR\_180x152} was run on an active region at 10:22~UT. This obtained spectra with the 2\as\ slit over a 180\as\ $\times$ 152\as\ area with a 6\as\ step size. A 60\as\ $\times$ 60\as\ block at the center of the active region was averaged to yield an intensity of 1182~\ecss\ with the \citet{2014ApJS..213...11W} calibration. Scaling with the nearest-in-time AIA 211~\AA\ image then gives an estimated full-disk intensity of 158~\ecss. This is only 14\%\ larger than the intensity obtained from HOP 130, remarkably similar to the result for \ion{Fe}{xii} \lam195.12.
}

%-------------------
\section{Annulus size}\label{app.annulus}

The annulus intensity used to obtain the short-range scattered light component was chosen to have radii between 30\as\ and 50\as. The inner boundary was set by the size of the Venus shadow.  The outer boundary was set with a view to apply the same method to EIS data, which generally has small fields of view. Given that the scattered light formula for AIA is recommended for use with any AIA data, we address what effect different radii would make. 

\begin{figure}[t]
\plotone{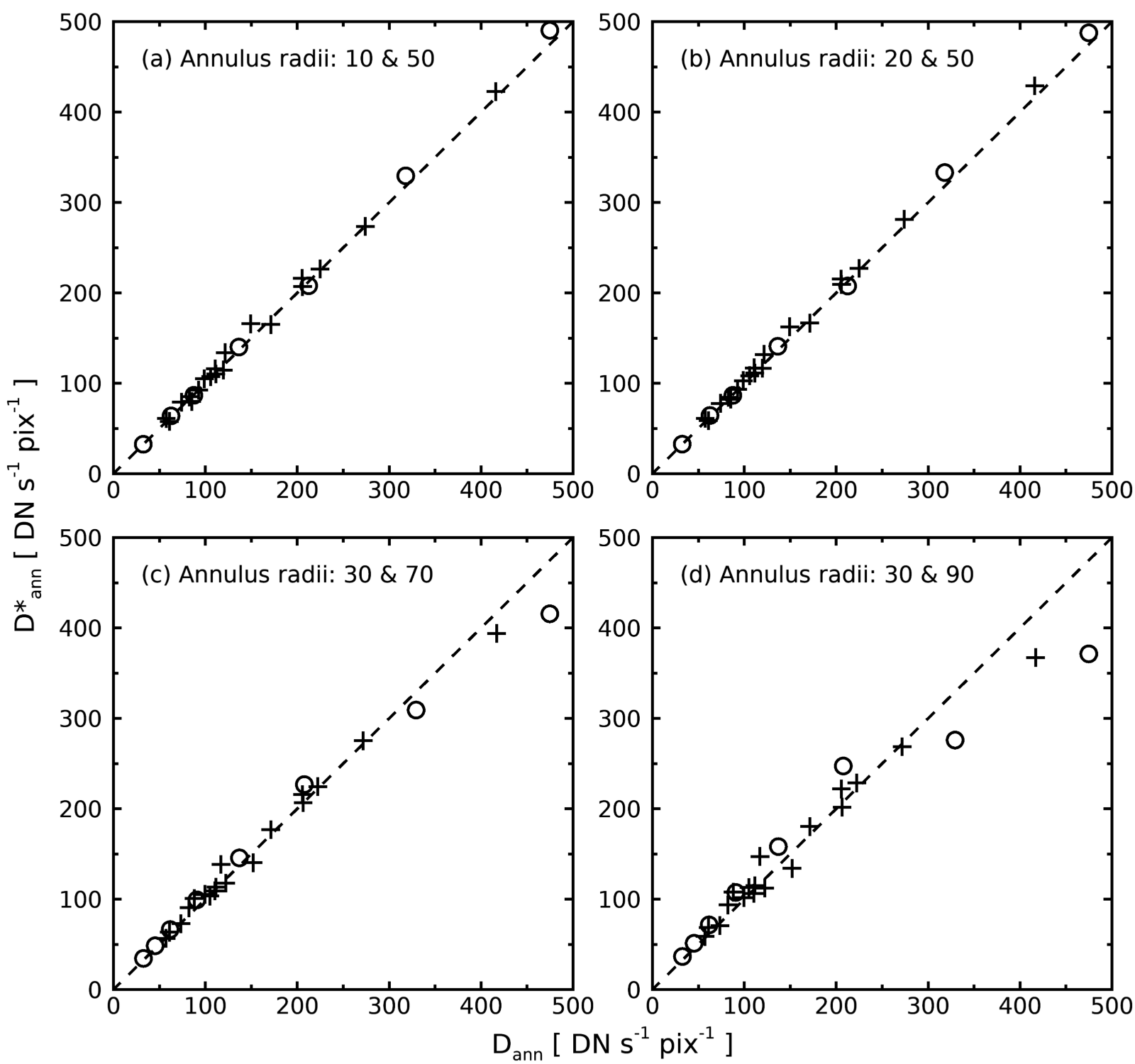}
\caption{Comparisons of AIA annuli intensities for alternative choices of the annuli radii ($D^*_\mathrm{ann}$), with intensities for the original choice of annulus radii ($D_\mathrm{ann}$). Panels (a) and (b) show the results for reducing the inner radius to 10\as\ and 20\as. Panels (c) and (d) show the results for increasing the outer radius to 70\as\ and 90\as. Crosses/circles represent locations where Venus was inside/outside the solar limb.}
\label{fig.aia-rad}
\end{figure}

In Figure~\ref{fig.aia-rad} we show how the AIA 193~\AA\ intensity averaged over the annulus during the Venus transit varies with the size of the annulus. Only the 20~min cadence data were used and so the additional frames mentioned in Section~\ref{sect.aia} were neglected. The X-axis for each plot is the annulus intensity measured in Section~\ref{sect.aia}, i.e., using inner and outer radii of 30\as\ and 50\as. The upper two panels show the effect of reducing the inner radius to 10\as\ and 20\as. Since the Venus radius is 30\as, these intensities were computed using the AIA image frame 20~min prior to Venus reaching that location. For example, at 00:20~UT the center of Venus was at ($-134,615$), so the annulus centered at this location but at time 00:00~UT was used. (For the first Venus location, the annulus intensity for the following image frame was used.)  The two lower panels show the effect of increasing the outer radius to 70\as\ and 90\as. (The annuli were centered on Venus for these cases since the inner radius remained at 30\as.)

As can be seen, the agreement between the annuli intensities is excellent. The largest differences compared to the original annulus intensities are 11\%, 9\%, 18\%, and 26\%\ for plots (a), (b), (c) and (d), respectively. This demonstrates that the choice of radii 30\as\ and 50\as\ for the annulus radii is a reasonable one for estimating the short-range scattered light.

%-------------------------------
\section{The Effect of a Nearby, Bright Active Region}\label{app.ar-model}

\pry{A potential limitation of our scattered light model is the extreme scenario of a bright active region close to the region of interest, but outside of the 50\as\ outer radius of the annulus used for estimating the short-range scattered light component. In this section we model this scenario.}

\begin{figure}[t]
\centering
    \includegraphics[width=\textwidth]{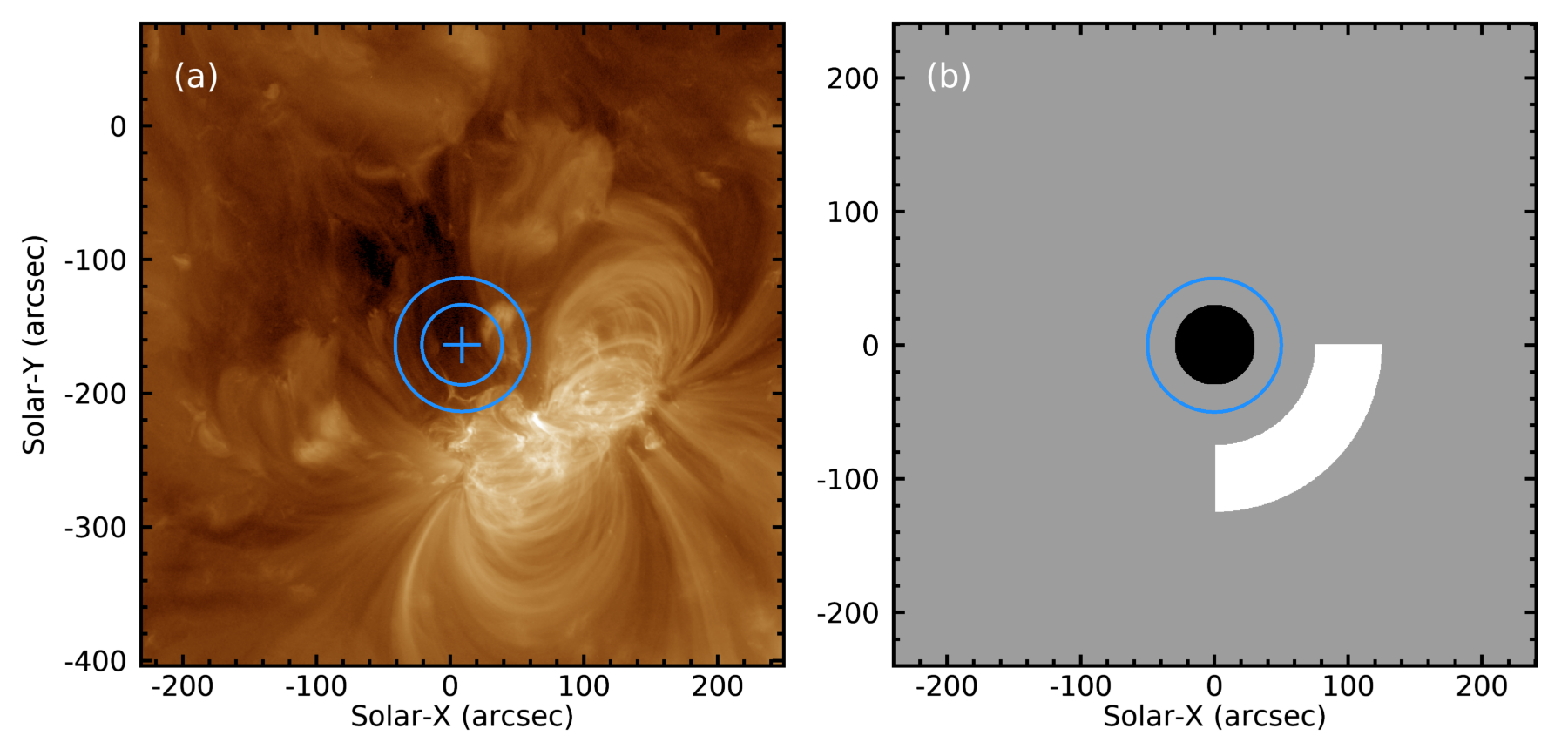}
\caption{(a) an AIA 193~\AA\ image with a logarithmic intensity scaling showing AR 11158. A \textit{blue cross} marks the location at which the scattered light is estimated, and the \textit{concentric circles} identify the annulus region used for assessing scattered light. (b) a synthetic intensity image for modeling scattered light. \textit{Black} indicates a region of zero intensity, \textit{gray} indicates quiet Sun emission, and \textit{white} active region emission (see main text for details).}
\label{fig.ar-model}
\end{figure}

\pry{We create a model based on a solar image from 2011 February 14 at 10:55~UT. Active region AR 11158 was close to disk center (Figure~\ref{fig.ar-model}a), and we consider the case of determining the scattered light in a dark region, indicated by a \textit{blue cross}, to the north-east of the active region center. The brightest part of the active region lies outside of the annulus (\textit{blue circles}). AR 11158 was very active at this time, producing six C-class flares on February 14 and an X-class flare on February 15. The time 10:55~UT was chosen as it corresponded with a lull in the flaring activity. The intensity at the location of the \textit{blue cross}, averaged over a $5\times 5$ pixel block is 42.1~DN~s$^{-1}$~pix$^{-1}$. Applying the deconvolution algorithm of \citet{aia-psf} gives an intensity at the same location of 23.3~DN~s$^{-1}$~pix$^{-1}$, and thus the scattered light is 18.8~DN~s$^{-1}$~pix$^{-1}$. Since the \citet{aia-psf} PSF does not account for long-range scattered light \citep{2020SoPh..295....6S} then it is reasonable to compare this value with the short-range component derived from our formula (Equation~\ref{eq.aia}). We find 13.3~DN~s$^{-1}$~pix$^{-1}$, and thus we consider the active region to enhance the scattered light by 41\%.}

\pry{As a sanity check to confirm that the AR is responsible for the enhanced scattered light, we consider the idealized case shown in Figure~\ref{fig.ar-model}(b). A circle of radius 30\as\ in the center of the image has zero intensity and the surroundings have a uniform intensity of 123~DN~s$^{-1}$~pix$^{-1}$ that was obtained by selecting a quiet region patch in panel (a). The white, quarter-circle region is placed between radii 75\as\ and 125\as\ from the image center, and mimics the active region in panel (a). The intensity is set to 1651~DN~s$^{-1}$~pix$^{-1}$, corresponding to the average intensity of the active region core.}

\pry{This idealized image scene is convolved with the \citet{aia-psf} PSF, and the resulting intensity at the center of the black circle is found to be 16.5~DN~s$^{-1}$~pix$^{-1}$. Setting the ``active region" to the quiet region intensity and performing the convolution with the PSF yields an intensity of 11.1~DN~s$^{-1}$~pix$^{-1}$. These values are quite consistent with the deconvolution procedure performed on the actual image scene, and the short-range scattered light component derived from our Equation~\ref{eq.aia}. This gives confidence that the active region is responsible for the additional scattered light.}

\pry{In summary, our short-range scattered light formula based on an annulus of outer radius 50\as\ will underestimate the scattered light by around 50\%\ if there is an intense active region nearby but outside of this radius. For a fainter active region and/or one positioned further away from the region of interest, this effect can be expected to be much less.}

%-------------------------------------------------
% \section{Wavelength dependence of scattered light}

% Sect.~\ref{sect.summary} suggests that the EIS scattered light formula can be applied to other emission lines formed at temperatures close $\log\,T=6.15$, including lines of \ion{S}{x} and \ion{Si}{x} that are found at wavelengths 258--265~\AA, significantly longer than \ion{Fe}{xii} \lam195.12. Would scattered light be expected to show the same behavior at these longer wavelengths?

% Figure~\ref{fig.193-304} reproduces the AIA 193~\AA\ scattered light curve for the 30\arcsec\ to 50\arcsec\ annulus from Figure~\ref{fig.aia-ann}. Overplotted is the curve obtained from the AIA 304~\AA\ PSF, and average intensity within $r\le 5$\arcsec\ is 0.045 and 0.049 for the 193~\AA\ and 304~\AA, respectively. The filter used for the 304~\AA\ channel is very similar to that used for the 193~\AA\ channel \citep{aia-psf} and so the differences are largely due to the wavelength dependence.

% An alternative is to use the 2017 September 10 X-flare, which showed a beautiful diffraction pattern in EIS that, since observed above the limb, is clearly seen in the data. A

% \begin{figure}[t]
% \epsscale{0.7}
% \plotone{plot_annulus_193_304_comparison.png}
% \caption{.}
% \label{fig.193-304}
% \end{figure}

\end{document}